\documentclass[journal]{IEEEtran}
\ifCLASSINFOpdf
\else
\fi

\usepackage{comment}
\usepackage{amsmath,amssymb,amsfonts}
\usepackage{algorithmicx}
\usepackage[ruled,linesnumbered]{algorithm2e}
\usepackage{graphicx}
\usepackage{textcomp}
\usepackage{xcolor}
\usepackage{float}
\usepackage{pifont}
\usepackage{diagbox}
\usepackage{setspace} 
\usepackage{multirow}  
\usepackage{verbatim}
\usepackage{tikz} 
\usepackage{xcolor} 
\usepackage{threeparttable}
\usepackage{caption}
\usepackage{ragged2e}
\usepackage{subfigure}
\usepackage{ulem}
\usepackage{hyperref}

\begin{document}
% \title{One-Step write: Improving the endurance of MLC STT-MRAM cache with compression}
%\title{LHMcache: A \underline{L}ow Latency and \underline{H}igh Endurance \underline{M}LC STT-MRAM based Cache System}
\title{CMD: A \underline{C}ache-assisted GPU \underline{M}emory \underline{D}eduplication Architecture}

\author{Wei Zhao, Dan Feng,~\IEEEmembership{IEEE Fellow}, Wei Tong, Xueliang Wei, Bing Wu
\thanks{The authors are with the Wuhan National Laboratory for Optoelectronics,
	Key Laboratory of Information Storage System, Engineering Research Center
	of data storage systems and Technology, Ministry of Education of China
	(School of Computer Science and Technology, Huazhong University of Science
	and Technology). E-mail: 1996hustzw@gmail.com, (dfeng, Tongwei, wubin200).}% <-this % stops a space
\thanks{Dan Feng and Wei Tong are the corresponding authors. E-mail: (dfeng,
	Tongwei@hust.edu.cn).}% <-this % stops a space
\thanks{This work was sponsored in part by the National Natural Science
	Foundation of China under Grant 61832007, Grant 61821003, Grant 62202195 and Grant
	62172178.}}


\markboth{IEEE TRANSACTIONS ON COMPUTER-AIDED DESIGN OF INTEGRATED CIRCUITS AND SYSTEMS}%
{Shell \MakeLowercase{\textit{et al.}}: Bare Demo of IEEEtran.cls for IEEE Journals}



\maketitle

\begin{abstract}
	Massive off-chip accesses in GPUs are the main performance bottleneck, and we divided these accesses into three types: (1) Write, (2) Data-Read, and (3) Read-Only. Besides, We find that many writes are duplicate, and the duplication can be \texttt{inter-dup} and \texttt{intra-dup}. While \texttt{inter-dup} means different memory blocks are identical, and \texttt{intra-dup} means all the 4B elements in a line are the same. In this work, we propose a cache-assisted GPU memory deduplication architecture named \textbf{CMD} to reduce the off-chip accesses via utilizing the data duplication in GPU applications. CMD includes three key design contributions which aim to reduce the three kinds of accesses: (1) A novel GPU memory deduplication architecture that removes the \texttt{inter-dup} and \texttt{inter-dup} lines. As for the \texttt{inter-dup} detection, we reduce the extra read requests caused by the traditional read-verify hash process. Besides, we design several techniques to manage duplicate blocks. (2) We propose a cache-assisted read scheme to reduce the reads to duplicate data. When an L2 cache miss wants to read the duplicate block, if the reference block has been fetched to L2 and it is clean, we can copy it to the L2 missed block without accessing off-chip DRAM. As for the reads to \texttt{intra-dup} data, CMD uses the on-chip metadata cache to get the data. (3) When a cache line is evicted, the clean sectors in the line are invalidated while the dirty sectors are written back. However, most read-only victims are re-referenced from DRAM more than twice. Therefore, we add a full-associate FIFO to accommodate the read-only (it is also clean) victims to reduce the re-reference counts. Experiments show that CMD can decrease the off-chip accesses by 31.01\%, reduce the energy by 32.78\% and improve performance by 37.79\%. Besides, CMD can improve the performance of memory-intensive workloads by 50.18\%.
\end{abstract}



\maketitle

\IEEEdisplaynontitleabstractindextext

% \fi
\IEEEpeerreviewmaketitle

\vspace{20pt}

\IEEEraisesectionheading{\section{Introduction}\label{sec:introduction}}
Due to GPUs' massive computational parallelism, modern Graphics Processing Units (GPUs) are widely used in deep learning, graph processing, and scientific programs. In the GPU micro-architecture, GPUs use streaming multiprocessors (SMs) to run many parallel application threads. To satisfy the demand of quickly rising GPU computational applications, commercial GPUs scale the number of SMs and increase the memory bandwidth of the on-chip cache and off-chip DRAM. For example, NVIDIA scaled memory bandwidth from 177 GB/s to 900 GB/s while extending the number of SMs from 14 in Fermi to 80 in Volta~\cite{zhao2019adaptive}.

NVIDIA GPUs typically use a two-level cache structure. Due to chip area limitation, the GPU L2 cache is smaller than the CPU's last-level cache. For example, the L2 cache in Tesla V100 GPU is 6MB, while the last-level cache in a mainstream server CPU can reach 256MB~\cite{zhao2022low}. The extremely limited on-chip memory leads to a lot of off-chip access, which is the main performance bottleneck. Previous work FUSE~\cite{zhang2019fuse} showed that nearly 75\% of the execution time and 70\% of the energy consumption was used to access off-chip DRAM. The massive off-chip DRAM accesses become the bottlenecks of GPU. Therefore, current GPU caches are organized as sectors, and only the dirty sectors are written back to reduce the off-chip bandwidth pressure.

Several cache-side and memory-side techniques are proposed to reduce the off-chip accesses and relieve the DRAM bandwidth pressure. For example, in the cache optimizations, FUSE~\cite{zhang2019fuse} and DeepNVM++~\cite{inci2021deepnvm++} replace the SRAM-based cache with the emerging STT-MRAM technology to increase the cache capacity. Moreover, Morpheus~\cite{darabi2022morpheus} extends the L2 cache capacity in GPU systems by using idle GPU core resources.
Besides, some memory-side works consider reducing the DRAM bandwidth pressure by reducing the transferred data between the DRAM and L2 cache. Some works propose effective compression techniques for GPU to reduce the data bandwidth pressure and memory footprint~\cite{sathish2012lossless, kim2016bit, rhu2018compressing, choukse2020buddy}. In addition, several CPU memory deduplication and compression techniques~\cite{zuo2018improving, park2021bcd} can reduce the outgoing requests or the CPU memory footprint. However, the CPU memory deduplication cannot be directly used in the GPU. Therefore, we mainly focus on designing the GPU memory deduplication architecture to improve the performance, and our work is orthogonal to the on-chip optimization and compression techniques. To implement the GPU memory deduplication, there are two challenges:
\textit{\textbf{(1) How to remove the duplicate blocks?}}
\textit{\textbf{(2) How to manage the duplicate blocks?}} Firstly, in CPU memory deduplication, They usually use a weak hashed value to detect duplicate data. If the hash value is matched, the two blocks are the same. But this way leads to extra reads in the detection process, which is used to avoid the hash collision. The extra read requests in GPU will degrade the performance. Secondly, the management granularity in the CPU usually is a cache line, i.e., 64B. However, the GPU sector cache varies the written back size from 32B to 128B. Therefore, how to manage these blocks is also a challenge.

In this paper, we propose our \textbf{C}ache-assisted GPU \textbf{M}emory \textbf{D}eduplication (\textbf{CMD}) architecture. We first analyze the off-chip accesses, and we classify them into three categories: (1) Write, (2) Data-Read, and (3) Read-Only. We find that many writes are duplicate, and the duplications can be intra-line and inter-line. As for the \texttt{intra-dup}, the 4B data elements in a memory block are all the same, and for \texttt{inter-dup} lines, they are duplicated with other lines. CMD can reduce these duplicate writes and solves the challenges above. CMD utilizes the feature that GPU is not sensitive to latency, so CMD adopts the strong hash way to avoid the extra reads for the hash collision check. Next, we propose several metadata techniques to manage duplicate blocks. Besides removing the duplicate write requests, as for the read requests that access the duplicate blocks (we name them duplicate reads), we propose a cache-assisted read technique to reduce the off-chip duplicate reads. Moreover, many read-only blocks are re-accessed from DRAM several times. Therefore, we use a small read-only FIFO buffer to store the read-only sector victims to avoid the re-accesses. CMD presents three key techniques to reduce the three kinds of off-chip accesses we mentioned above, and the main design contributions are as follows:

\begin{itemize}
	\item \textit{\textbf{Efficient GPU Memory Write Deduplication.}} To the best of our knowledge, our work is the first work that proposes the memory deduplication architecture for the GPU. We present intra-line and inter-line deduplication. As for the \texttt{intra-dup} lines, the 4B data is co-located in the address mapping cache. As for \texttt{inter-dup} lines, we use the strong hash method to avoid the extra read-verify processes. Besides, we propose several metadata techniques to manage the duplicate blocks.
	
	\item \textit{\textbf{Cache-Assisted Read to Reduce Duplicate Reads.}} We propose to reduce the duplicate reads with the assistance of the L2 cache and metadata cache. As for \texttt{intra-dup} blocks, its data is co-located in the address mapping cache. Once it is hit in the metadata cache, the data is returned. For \texttt{inter-dup} reads, the reference block may have been fetched into the L2 cache. Once the reference block is clean, we can copy it into the L2 missed block and avoid the off-chip accesses.
	
	\item \textit{\textbf{Read-Only FIFO for Clean Victims.}} When an L2 cache line is selected as a victim, the dirty sectors are written back to DRAM, while clean sectors are invalidated. However, these clean sectors will be re-referenced several times. Hence, we set a FIFO buffer to accommodate these read-only (also clean) sector victims and reduce the re-accesses to DRAM.
	
	\item \textit{\textbf{System-Level Implementation and Evaluation.}} We implement our scheme in GPGPU-sim 4.0~\cite{bakhoda2009analyzing,khairy2020accel}. Experiments show that CMD can decrease the off-chip access by 31.01\%, reduce the energy by 32.78\% and improve performance by 37.79\%. Besides, CMD can improve the performance of memory-intensive workloads by 50.18\%.
\end{itemize}

\section{Background}
\begin{comment}
	\subsection{The basics of DNN}
	DNN is composed of several layers. Fig. shows the structure of Convolution Neural Network (CNN). 
\end{comment}
\subsection{The Memory Hierarchy of GPU}
\begin{figure}[htbp]
	\vspace{-5pt}
	\centerline{\includegraphics [width=0.45\textwidth]{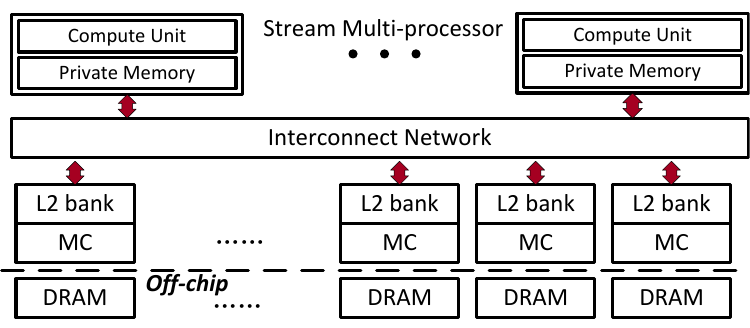}}
	\vspace{-5pt}
	\caption{The memory architecture of GPU.}
	\label{memory}
	\vspace{-5pt}
\end{figure}

Figure \ref{memory} shows the memory architecture of GPU that employs several streaming multi-processors (SM). Each SM contains a computing unit and private on-chip memories. Usually,
the computing unit has 32 CUDA cores, which can operate a group of 32 independent threads called \textit{warp} in a GPU clock cycle. To keep the pipeline busy, GPU schedules hundreds of threads within each SM and supports seamless context switches at the hardware level to eliminate the context switch penalty. The on-chip memories can cache the frequently used data and instructions, quickly responding to SM requests. Due to the limited area for private on-chip memories, several SMs share the L2 cache, while the L2 cache has several memory partitions. The memory controller manages the read/write requests between the on-chip memory and off-chip DRAM in each memory partition unit. While the interconnect network provides the topology routines for the SMs and memory partitions, which enables SMs to access all the memory partitions.
\subsection{GPU Sector Caches}
As for GPU, the cache line size is usually 128B, and each cache line is divided into four 32B sectors. Therefore, the requested granularity for the GPU cache is a sector~\cite{sector_cache}. While an L2 cache line victim is written back to the off-chip DRAM, only the modified sectors are written back. In this way, the demand for DRAM bandwidth can be decreased. When the L2 cache encounters a cache miss, a sector in the DRAM is fetched to the L2 cache. Therefore, for the main memory in GPU, the write request size varies from 32B to 128B, while the read request size is 32B in the sectored GPU cache.
\section{Motivation}
Due to the limited L2 cache in GPU architecture, many requests need to access the DRAM. However, the DRAM accesses are inefficient, significantly increasing latency and energy consumption. From the evaluations in previous studies, the latency consumed by outgoing memory requests, on average, accounts for 75\% of the total GPU device execution time. Similarly, the I/O requests for off-chip data accesses take up 71\% of the total GPU energy~\cite{zhang2019fuse}. In this paper, we analyze the characteristics of off-chip requests and avoid as many off-chip accesses as possible. 

\begin{figure}[htbp]
	\vspace{-5pt}
	\centerline{\includegraphics [width=0.5\textwidth]{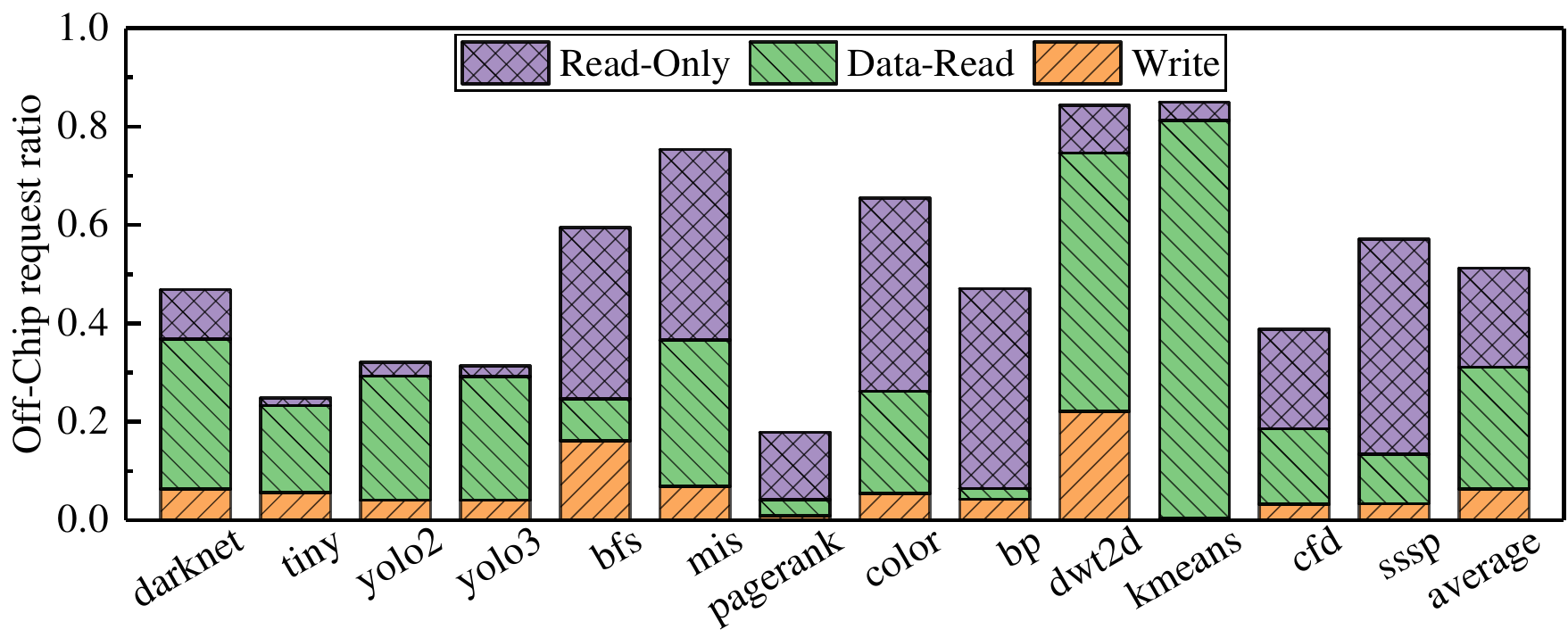}}
	\vspace{-5pt}
	\caption{The off-chip access ratio and requests breakdown.}
	\label{req_ana}
	\vspace{-5pt}
\end{figure}

\textbf{\textit{1) The Breakdown of Massive Off-Chip Accesses.}}
Figure \ref{req_ana} shows the breakdown of off-chip accesses. Overall, 51.21\% access is outgoing to DRAM with the system configuration in TABLE \ref{system_config}. Besides, we classify the off-chip accesses into three categories: (1) Write Requests; (2) Data-Read Requests, which read the blocks that have been written back from the L2 cache; (3) Read-only Requests, which access the read-only data in the application, such as the weights data in the DNN inference program. In detail, 6.38\% accesses are \texttt{Write} requests, 24.75\% are \texttt{Data-Read}, and 20.08\% are \texttt{Read-Only} requests. This paper proposes several techniques to reduce the three kinds of memory requests to improve GPU performance and reduce energy consumption. 

\begin{figure}[htbp]
	\vspace{-5pt}
	\centerline{\includegraphics [width=0.5\textwidth]{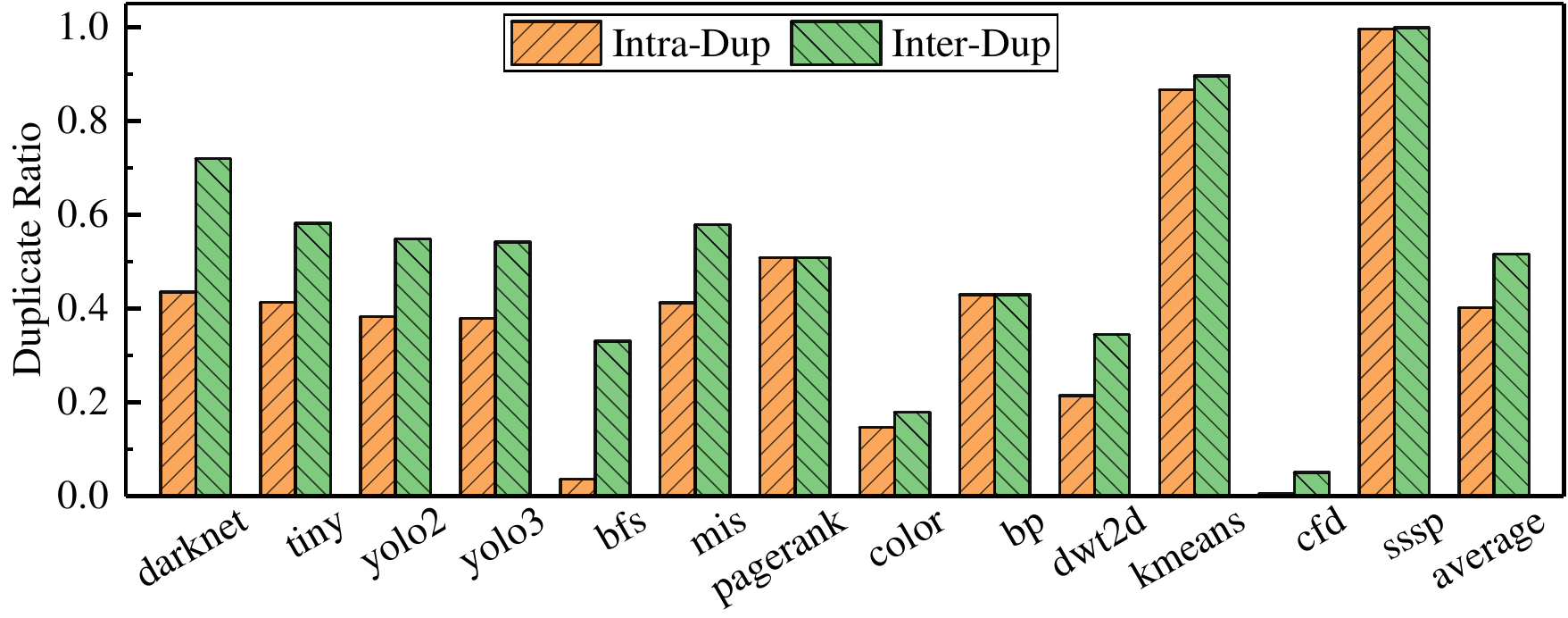}}
	\vspace{-5pt}
	\caption{The duplication ratio of several benchmarks.}
	\label{duplication}
	\vspace{-5pt}
\end{figure}

\textbf{\textit{2) Duplicate Writes in GPU.}}
We find massive writes in GPU are redundant. The redundancy includes two parts: (1) Intra-redundancy, i.e., all the data elements (4B granularity) are the same in one memory block. For example, one 128B block stores 32 \texttt{float} values, and all the values are the same. For this kind of memory block, we can classify them as \texttt{intra-dup} blocks. (2) Inter-redundancy. The memory blocks with different block addresses may be the same, and we call them \texttt{inter-dup} blocks. From the test results in Figure \ref{duplication}, we know 40.18\% of the blocks are intra-duplicate, and 51.58\% of the blocks are inter-duplicate.

\section{Design}
\begin{figure}[htbp]
	\vspace{-5pt}
	\centerline{\includegraphics [width=0.45\textwidth]{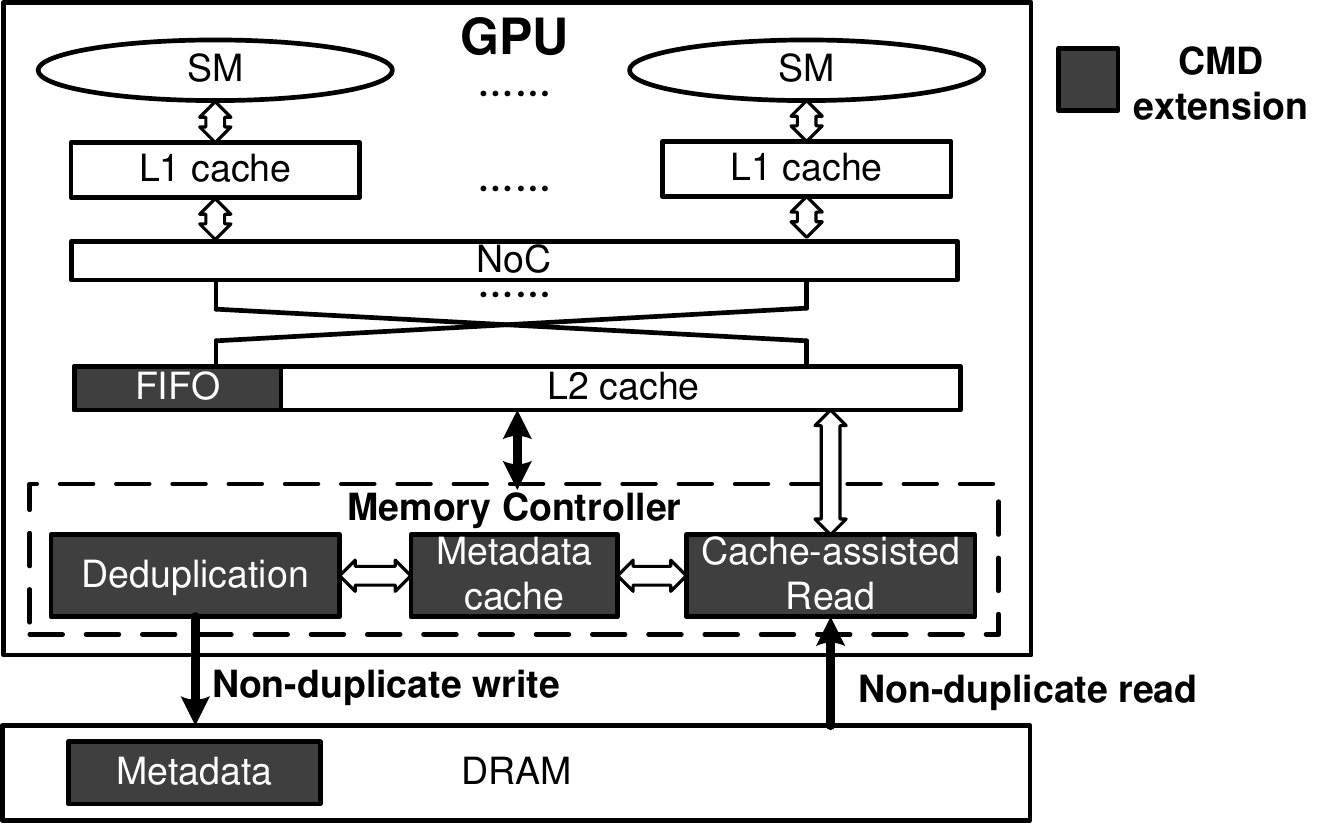}}
	\caption{The overview of our CMD architecture.}
	\label{overall}
	\vspace{-5pt}
\end{figure}

Figure \ref{overall} shows the overall architecture of our Cache-assisted GPU Memory Deduplicated (CMD) system. In CMD architecture, three main techniques aim to reduce the three kinds of requests in Figure \ref{req_ana}. Firstly, the deduplication logic detects the duplicate writes and maps several duplicate blocks to one physical block. Therefore, only the non-duplicate blocks are written to off-chip DRAM. Secondly, there are many duplicate reads as well. Therefore, we propose a cache-assisted read scheme to fetch data in the metadata cache or L2 cache, and the non-duplicate data are read from the DRAM. Thirdly, we find that many clean read-only blocks are re-referenced from DRAM by several times. Hence, we propose a read-only First In First Out (FIFO) buffer to store these clean sector victims, not simply abandoning them. The FIFO is placed in the L2 cache.

\subsection{Detect Duplicate Writes}
From Figure \ref{duplication}, we can see that there are many duplicate memory writes, i.e., \texttt{intra-dup} and \texttt{inter-dup}. As for the \texttt{intra-dup} lines detection, we can detect if all the 4B data elements are identical. As for \texttt{inter-dup} lines detection, current techniques use the hash value of the block data to identify if this line is duplicate. The deduplication process follows these steps in current deduplicated storage/memory systems. \ding{182}: The deduplication logic calculates the hash value of the written content. \ding{183}: The hash store identifies if the hash is hit. While the hash value hits/misses, and this block is duplicate/non-duplicate. If the hash misses and the hash store has an empty entry, this hash is inserted into the hash store.

\begin{figure}[htbp]
	\vspace{-5pt}
	\centering
	\subfigure[Weak hash with read\&comparison.]{
		\begin{minipage}[t]{0.5\linewidth}
			\centering
			\includegraphics[width=1.6in]{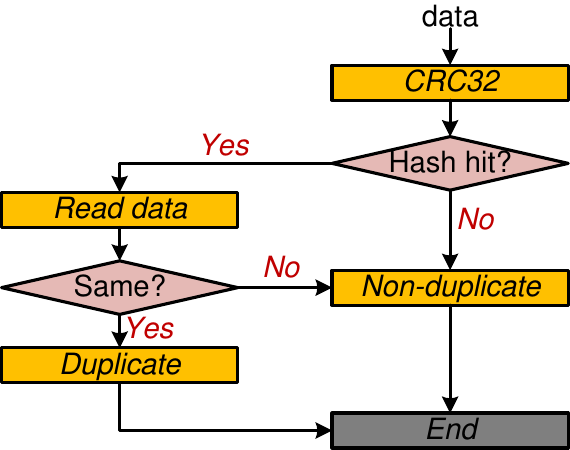}
			%\caption{fig1}
		\end{minipage}%
	}%
	\subfigure[Strong hash.]{
		\begin{minipage}[t]{0.5\linewidth}
			\centering
			\includegraphics[width=1.6in]{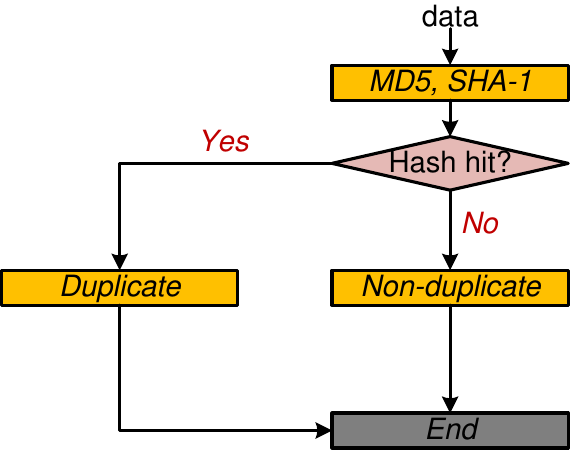}
			%\caption{fig2}
		\end{minipage}%
	}%
	\centering
	\vspace{-5pt}
	\caption{Two methods of duplicate memory lines detection.}
	\vspace{-5pt}
	\label{detect}
\end{figure}

\textbf{\textit{Hash Collision Discussion.}}
Hash logic transfers the 32B-128B data to smaller-size data like 16B in MD5. Therefore, two blocks may have the same hash value, i.e., hash collision. Figure \ref{detect} shows two methods for avoiding hash collision, which are commonly used in the deduplicated storage systems~\cite{xia2016comprehensive}. The first is (a) Read-Verify. When the calculated weak hash value is found in the hash store, a read-verify operation is used to identify if it is the same as the reference block. This way can remove the hash collision problem. But every deduplication process needs one read-verify operation, degrading overall performance. (b) Complex Hash is another alternative. Some storage systems, such as ZFS~\cite{rodeh2003zfs} and Dropbox~\cite{drago2012inside}, have considered using SHA-256 or MD5~\cite{rivest1992md5} hash algorithm to reduce the risk of hash collision probability. GPU system is bandwidth-sensitive and latency-insensitive. Extra read requests will degrade the system performance. Therefore, in our scheme, we adopt the scheme in Figure \ref{detect}(b).

\textbf{\textit{The Performance Discussion of two Hash Methods.}} The MD5 hardware implementation needs 228 \textit{cycles} (in SM core cycles) for 128B data~\cite{jarvinen2005hardware}, while simple hash like CRC needs 40 \textit{cycles}~\cite{huo2015high}. We implement three comparison schemes to investigate the performance improvement of the two hash methods. \texttt{ESD} is a state-of-the-art memory deduplication architecture in CPU memory that uses the weak hash method. ESD uses the Error Correction Code (ECC) of last level cache as the hash fingerprint, avoiding the hash calculation latency. \texttt{Dedup} is our method with strong hash, and \texttt{Dedup\_no\_latency} is the ideal \texttt{Dedup} without hash latency. The results show that \texttt{ESD} has a small performance loss due to extra read requests. \texttt{Dedup} can gain a 6.8\% IPC improvement, and it only loses 6.5\% IPC compared with the ideal \texttt{Dedup}. The reasons are: (1) GPUs are latency-insensitive and bandwidth-sensitive, and other operations can hide the long latency. (2) The write request is off the critical path of GPU DRAM. (3) We only compute the hash for \texttt{inter-dup} lines, and the \texttt{intra-dup} lines are bypassed. Therefore, the long-latency strong hash method is better for GPU memory deduplication.

\begin{figure}[htbp]
	\vspace{-5pt}
	\centerline{\includegraphics [width=0.48\textwidth]{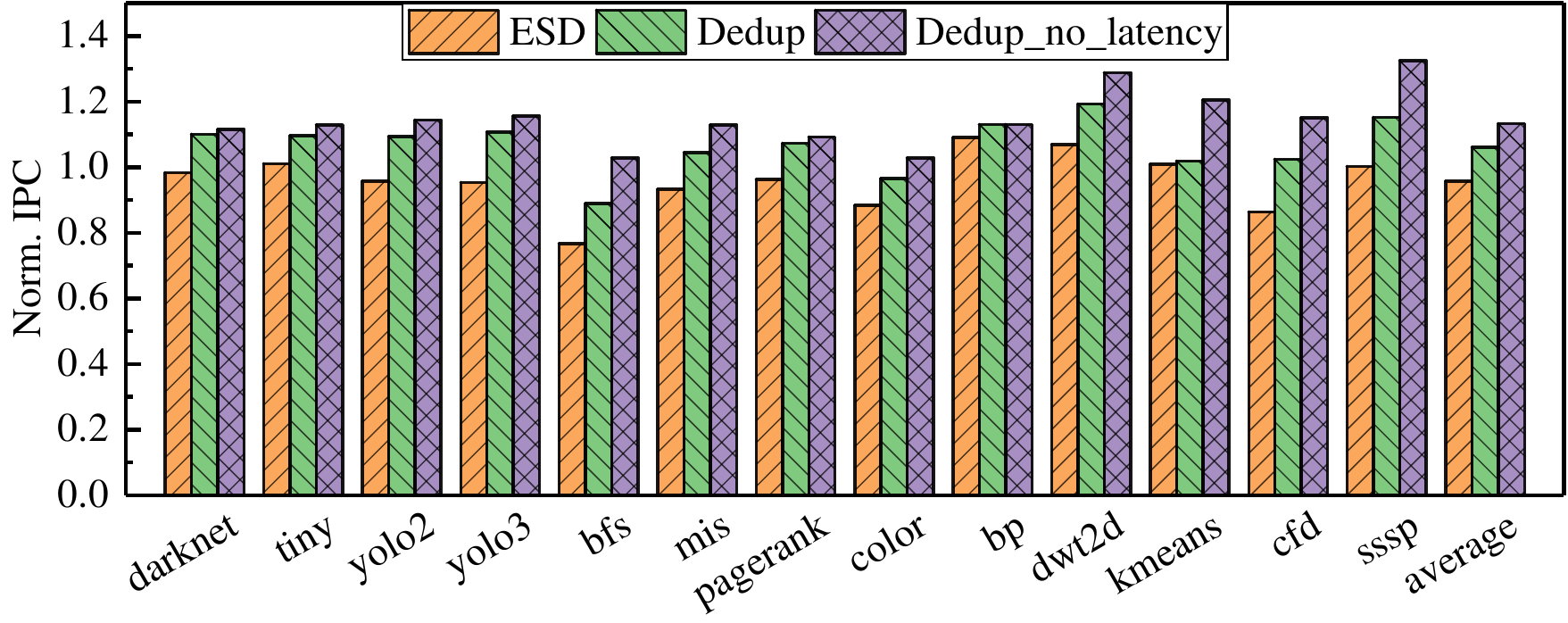}}
	
	\caption{The normalized performance of two hash methods.}
	\label{buffer_sen}
	\vspace{-5pt}
\end{figure}

\begin{figure}[htbp]
	\vspace{-5pt}
	\centering
	\subfigure[Need an extra read.]{
		\begin{minipage}[t]{0.5\linewidth}
			\centering
			\includegraphics[width=1.6in]{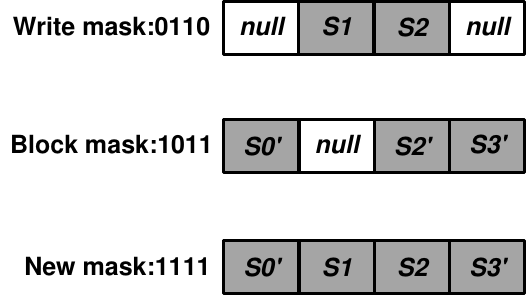}
			%\caption{fig1}
		\end{minipage}%
	}%
	\subfigure[No read.]{
		\begin{minipage}[t]{0.5\linewidth}
			\centering
			\includegraphics[width=1.6in]{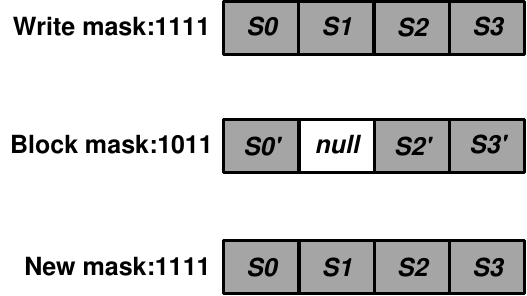}
			%\caption{fig2}
		\end{minipage}%
	}%
	\centering
	\vspace{-5pt}
	\caption{The two examples of sector coverage.}
	\vspace{-5pt}
	\label{sector_fig}
\end{figure}

\textbf{\textit{Sectors Coverage.}} The second challenge for GPU memory deduplication is the mutative written data size. In our deduplication system, we use the 128B cache line block as the management granularity. In comparison, we also explored sector-level deduplication, but this fine-grained way leads to high DRAM overhead. In our technique, the duplicate blocks' size can be 32B-128B, although the management granularity is 128B. When one block is mapped into a reference block, and it may be modified in the future writes, we should consider the sector mask coverage issue. Figure \ref{sector_fig} shows two examples. In Figure \ref{sector_fig}(a), the write mask is "0110", and the block's mask is "1011". In this case, we must read the block's data and compact it to form the new data, then use it to participate in the deduplication process. In comparison, in the case of Figure \ref{sector_fig}(b), the write mask can cover the block mask, which means all the old data in the block will be updated. Thus, we need no extra read. In summary, if the new mask and old mask satisfy the Equation 
\ref{condition}, we need not read the old data, and the mask is updated via Equation \ref{update}.

\begin{equation}
	new\_mask[i]>=old\_mask[i] \quad (i=0,1,2,3)
	\label{condition}
\end{equation}

\begin{equation}
	mask=old\_mask | new\_mask
	\label{update}
\end{equation}

\begin{comment}
	\begin{figure}[htbp]
		
		\centerline{\includegraphics [width=0.5\textwidth]{./fig/merge_buffer.pdf}}
		\vspace{-5pt}
		\caption{The read(non-blocking)/write process with a dedup hash buffer.}
		\label{merge_buffer}
		
	\end{figure}
\end{comment}

\begin{figure}[htbp]
	\vspace{-5pt}
	\centerline{\includegraphics [width=0.48\textwidth]{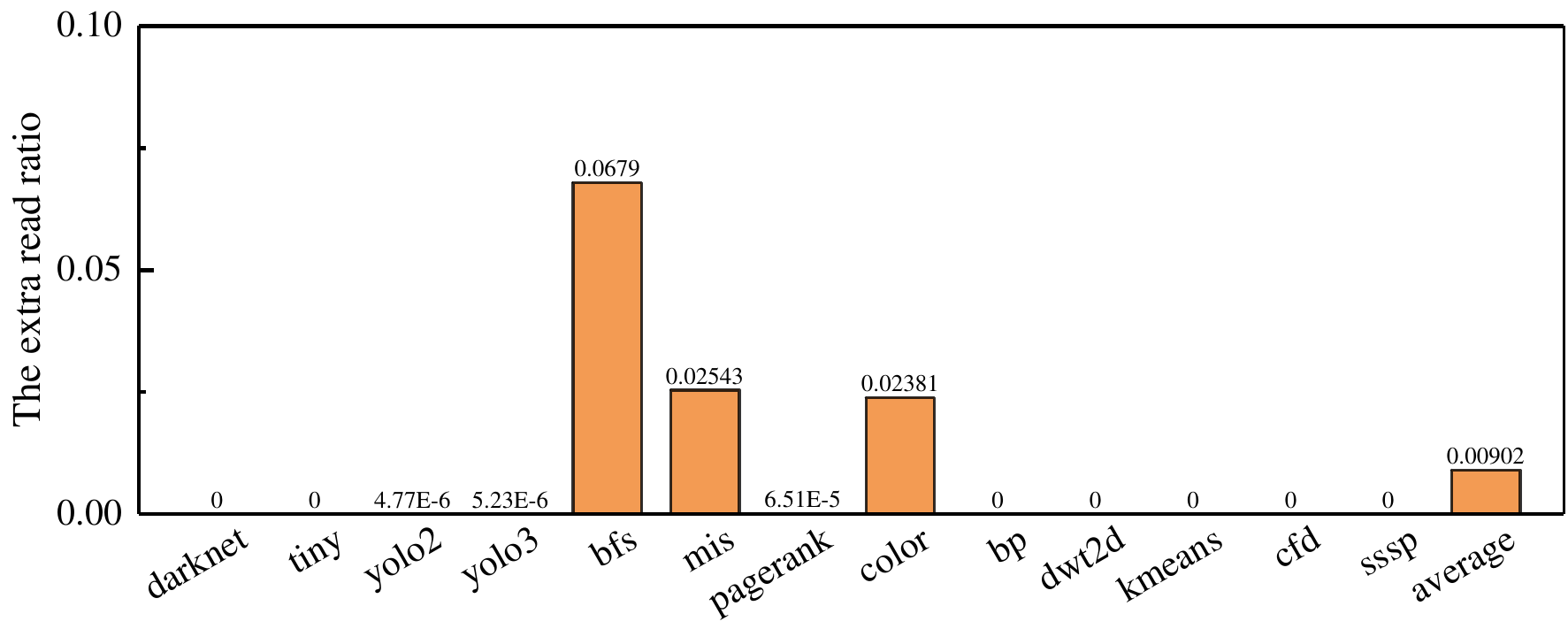}}
	\vspace{-5pt}
	\caption{The extra read request ratio.}
	\label{non_cover}
	\vspace{-5pt}
\end{figure}

Figure \ref{non_cover} shows the extra read requests in the write deduplication process. We can see that in most benchmarks like \texttt{darknet}, \texttt{tiny}, \texttt{pagerank} etc., the new sector mask can cover the old mask. This is because almost all their written data sizes are 128B, i.e., four sectors. Therefore we do not need to fetch the old block. As for \texttt{bfs}, \texttt{mis}, and \texttt{color}, the read ratio is below 7\%. Therefore, they suffer from little performance degradation, shown in Figure \ref{buffer_sen}. On average, the extra read ratio is 0.90\%.

\begin{figure*}[htbp]
	
	\centerline{\includegraphics [width=1.05\textwidth]{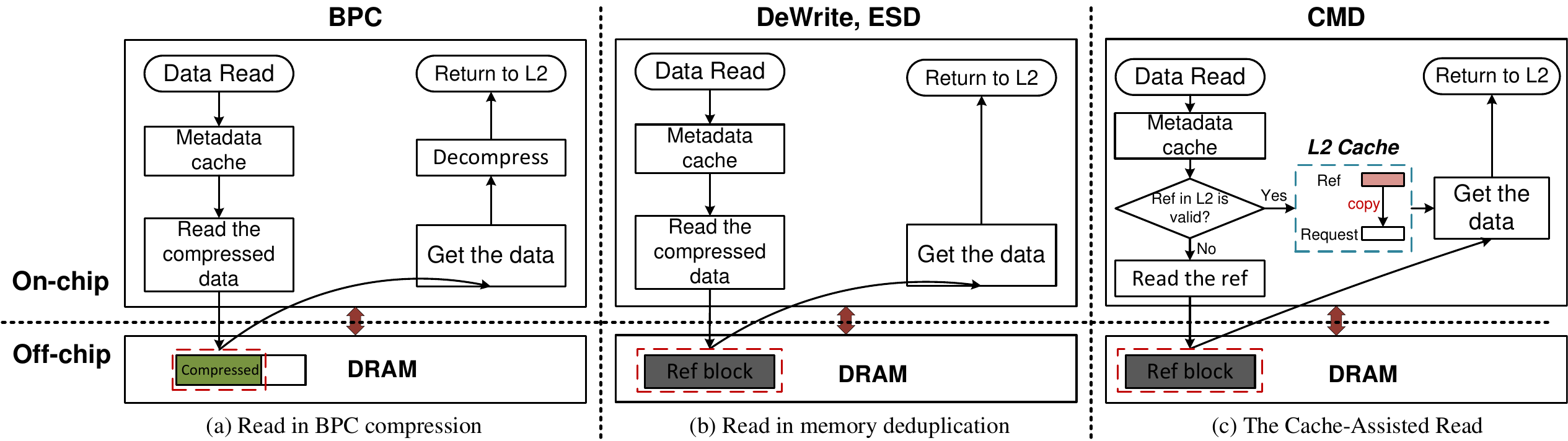}}
	\vspace{-5pt}
	\caption{The read procedures for the Data-Read requests.}
	\label{read_compare}
	\vspace{-5pt}
\end{figure*}
\subsection{Metadata Management}

We need metadata to record the block information and associate duplicate blocks to one reference block.

\textbf{\textit{Hash Store.}} The key idea of deduplication is to map several duplicate blocks to one reference block, and the on-chip hash bucket stores a bunch of hash entries of reference blocks. Each entry has three items, i.e., \texttt{[hash,ref\_addr,count]}. \texttt{hash} is the hash value of the reference block, and its size is 16B for the MD5 algorithm. \texttt{ref\_addr} is the reference block physical address, and its size is \texttt{4B}. While the address granularity is 128B, a 32-bit address can support 512GB memory in a DRAM memory partition, which is enough. \texttt{count} records the number of mapped blocks. If one block is the same as the reference block, \texttt{count} increases by one. Meanwhile, if one mapped block is changed, \texttt{count} decreases by one. We assign 2B for the \texttt{count} value. When \texttt{count} exceeds $2^{16}-1$, we classify this write as non-duplicate, while this case happens rarely.

\textbf{\textit{Address Mapping/Storing intra-duplicate data.}} As for deduplicated GPU memory, the mapping of the physical address and the data blocks are not one-to-one, and many addresses map to one data block. The address mapping table indicates the mapping relationship between duplicate blocks and the reference block, and each block has an entry. The data structure is like \texttt{[blk\_addr,ref\_addr]}. The address mapping table stores its referenced addresses \texttt{ref\_addr}, and the item size is \texttt{4B}. As for \texttt{intra-dup} blocks, its data is 4B, which can be stored in the address mapping table, and we use a simple flag to indicate what kind of data the entry stores.
\begin{comment}
	\textbf{\textit{Invert Hash Table.}} While one memory block which stores the base mapping address is rewritten, the old hash table entry is invalid. Therefore, the invert hash table is used to query and clean up the old entry in the hash table. Due to the hash table entry has one unique \texttt{ref\_addr} and \texttt{hash} pair, so after a rewrite process, we can query the hash store to cleanup the stale hash table entry according to the new inverted hash table. The data structure of inverted hash table entry is like \texttt{[blk\_addr,hash]}. Specially, due to we use a strong cryptographic secure hash function, the hash size is 128-bit, so we store the partial hash in the inverted hash table to reduce the capacity requirement. 
	
	\textbf{\textit{Address Mapping and Inverted Hash Tables Co-location.}} We notice that the entries of hash store and inverted hash table are similar. In each entry with \texttt{blk\_addr}, the corresponding value is either \texttt{hash} or \texttt{ref\_addr}, so we can store them together with 1-bit flag to indicate which kind of metadata is stored. For example, "0" indicates the entry stores the mapping address, while "1" is for partial hash.
\end{comment}

\textbf{\textit{Type Flag and Sector Mask.}} We use a 2-bit flag to indicate the data type of each block. There are four kinds of data. (1) Read-Only block; (2) Intra-duplicate block; (3) Intra-duplicate block; (4) Reference and non-duplicate block. The data structure is like \texttt{[blk\_addr,Flag]}. The default type flag is "00", once a block is modified, the type will be updated as well. The sector mask is similar, the data structure is like \texttt{[blk\_addr,Mask]}, and each 128B block equips 4-bit. When the referenced block is modified, we can use the type flag to find an empty intra-duplicate or inter-duplicate block, then map the data to the corresponding new block location.

CMD only equips the hash store in the on-chip memory controller, avoiding many off-chip hash entry requests. When the hash store is full, the corresponding LRU hash store entry, whose \texttt{count} is 1 (i.e., the duplicate block is the reference block), will be cleaned. If there is no hash entry with the \texttt{count} equals 1, this block is assumed as a non-duplicate block. Although this way may affect the deduplication ratio, the influence is not obvious. From the evaluation results, nearly 90\% duplicate writes can be removed. Besides, placing the hash store only on-chip can reduce the performance loss due to the off-chip requests for hash entry. As for the address mapping table, type flag, and sector mask, we keep the hot entries on-chip. When a metadata cache occurs a miss, CMD will fetch the metadata with 32B granularity.
\begin{comment}
	\textbf{\textit{The Replacement Policy of Metadata Cache.}} As for the metadata cache write, we use the LRU replacement policy to get the victim while the metadata cache is full. While for the read requests, while the metadata cache is miss, we send the evict the LRU victims and fetch several entries from memory to metadata cache with the assistance of cache spatial locality.
\end{comment}

\subsection{Cache-assisted Read Scheme}
To effectively mitigate the duplicate \texttt{Data-Read} requests, Figure \ref{read_compare} analyzes the read operations of \texttt{Inter-Dup} blocks using various methods. For GPU memory compression techniques like BPC ~\cite{kim2016bit} (Figure \ref{read_compare}(a)), during write operations, compression can reduce the amount of data being written. Consequently, during reads, it is only necessary to read the compressed data portion based on the compression status and compressed data address mapping in the metadata cache, thus reducing bandwidth consumption. For memory deduplication techniques in current CPU systems, such as DeWrite~\cite{zuo2018improving} and ESD~\cite{du2023esd} (Figure \ref{read_compare}(b)), the reference block address must first be obtained according to the metadata cache. Then the corresponding reference block data is fetched from DRAM and returned to the L2 cache. Therefore, traditional memory deduplication methods cannot reduce duplicate data reads, and they can only reduce the write requests and overall physical memory footprint. This approach is not friendly to workloads with many duplicate data read requests, as it cannot reduce duplicate data reads. Alternatively, caching frequently accessed reference block data in the memory controller can reduce duplicate reads. However, this method incurs significant on-chip cache overhead and substantially increases the memory controller's die area, as many data entries are required to achieve a high hit rate. Figure \ref{read_compare}(c) illustrates the \texttt{Inter-Dup} data read processes of the cache-assisted read (CAR) method. Due to the presence of temporal locality in GPUs, the corresponding duplicate block data will also be accessed when the reference block is fetched into the L2 cache. Therefore, when reading the duplicate block, if the reference block data is already in the L2 cache and valid, the reference block data in the L2 cache can be directly copied to the requested data block, thereby avoiding off-chip accesses for \texttt{Inter-Dup} read requests.

\begin{figure}[htbp]
	
	\centerline{\includegraphics [width=0.5\textwidth]{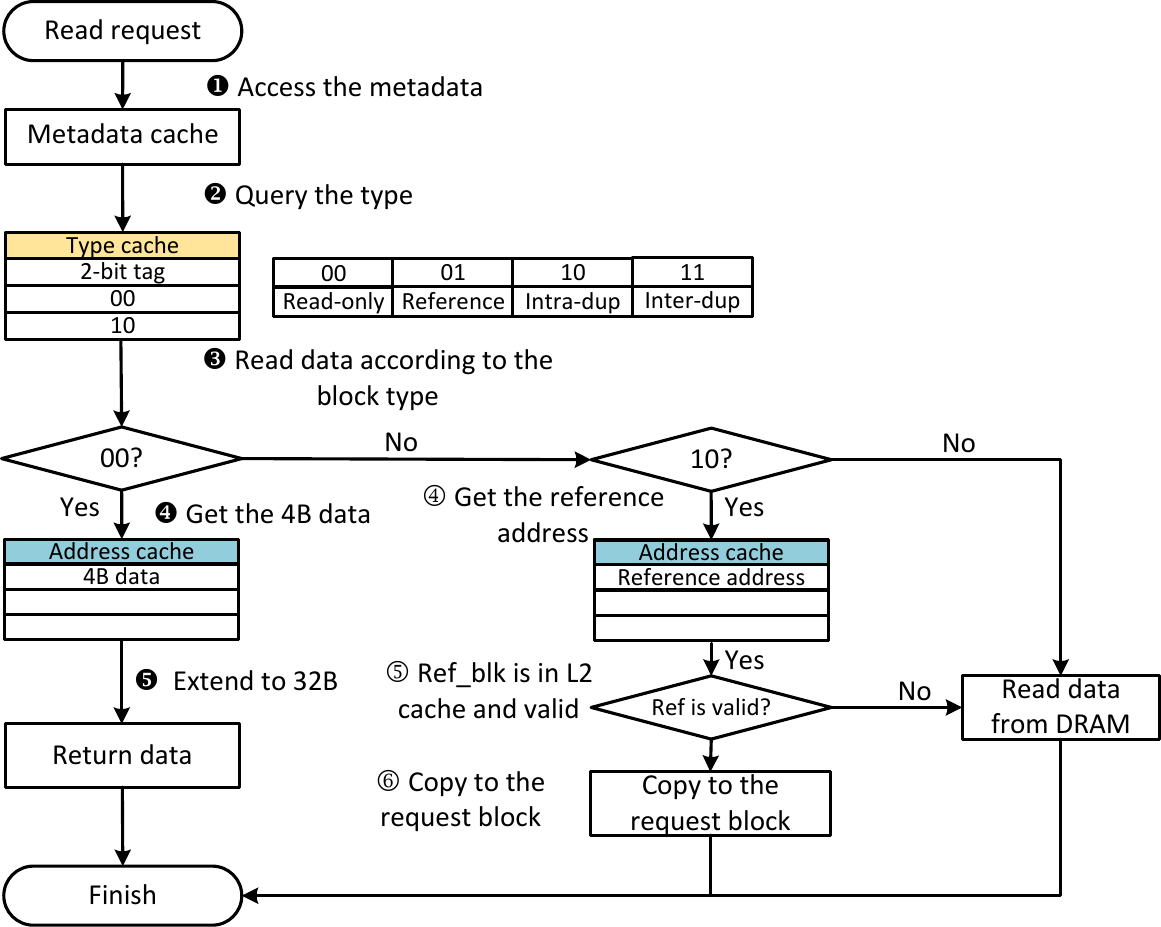}}
	\vspace{-5pt}
	\caption{The read procedures for the Data-Read requests.}
	\label{CAR}
	\vspace{-5pt}
\end{figure}

The Cache-Assisted Read (CAR) scheme utilizes metadata cache in the memory controller and L2 cache to assist the read operations, reducing \texttt{Data-Read} requests for both \texttt{inter-dup} and \texttt{intra-dup} blocks. Figure \ref{CAR} illustrates the data read process for the CAR scheme. When a \texttt{Data-Read} request arrives at the memory controller, the metadata cache is first accessed to determine whether the metadata is in the memory controller (Figure \ref{CAR}~\ding{182}). Subsequently, the type cache is queried to obtain the data block type (Figure \ref{CAR}~\ding{183}). Once the read type is known, different read operations are performed according to different types (Figure \ref{CAR}~\ding{184}). If the type flag is "01", it indicates that the requested block is \texttt{intra-dup}, and the address cache is accessed to obtain the stored 4-byte data (Figure \ref{CAR}~\ding{185}). Finally, the 4-byte data is extended to 32 bytes, and the data is returned to the upper-level cache to complete the read operation (Figure \ref{CAR}~\ding{186}). If the type flag is "10", it indicates that the block is inter-block duplicate data. First, the reference block address is obtained through the address mapping table (Figure \ref{CAR}~\ding{175}). Then, it is determined whether the reference block data already exists in the L2 cache (Figure \ref{CAR}~\ding{176}). Since DRAM data read requests are caused by L2 cache misses, if the reference block data is in the L2 cache and valid, the corresponding data can be directly copied to the requested data block in the L2 cache (Figure \ref{CAR}~\ding{177}) to eliminate off-chip DRAM access. If the type flag is "00" or "11", it indicates that the block is a read-only block or a reference/non-duplicate block. In this case, data can only be read from DRAM. Through the CAR read mechanism, a large number of off-chip reads for duplicate data can be eliminated.

\subsection{Read-Only FIFO for Clean Victims}
\begin{figure}[htbp]
	
	\centerline{\includegraphics [width=0.5\textwidth]{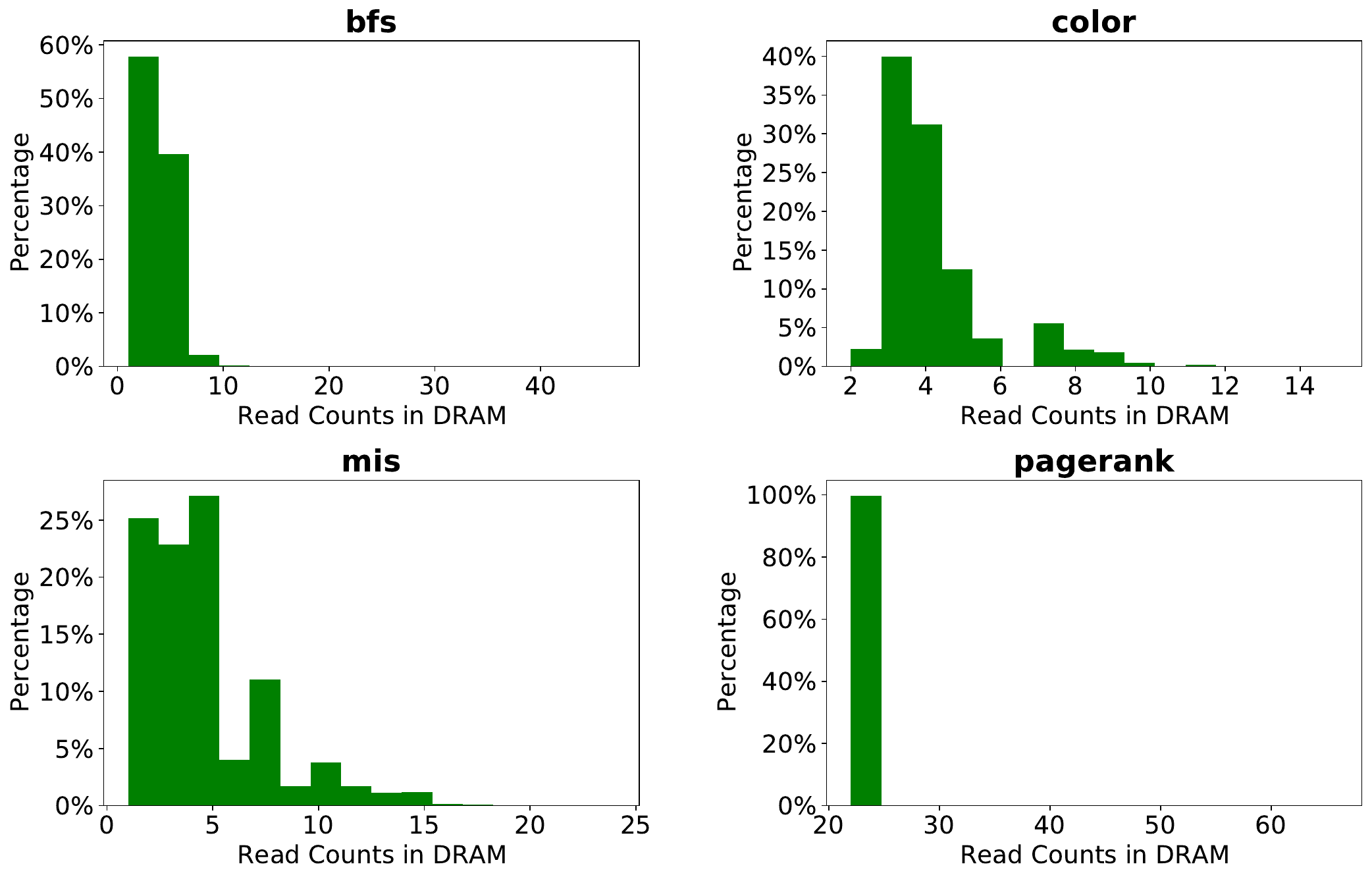}}
	\vspace{-5pt}
	\caption{The read counts distribution of read-only requests.}
	\label{read_only_hist}
	\vspace{-5pt}
\end{figure}

In addition to a large number of \texttt{Data-Read} requests, there are also many \texttt{Read-Only} requests. When read-only data is fetched from DRAM to on-chip, due to the limited capacity of the L2 cache in GPUs, the read-only blocks will be evicted by the replacement algorithm after a certain period of time. Subsequently, if the application requires the read-only data block multiple times, it needs to be read from DRAM again. As a result, read-only data is read multiple times from DRAM, which can lead to a large amount of off-chip accesses and degrade system performance. To verify this hypothesis, Figure \ref{read_only_hist} presents a histogram of the distribution of read-only block read counts from DRAM. It can be observed that in many application workloads, numerous read-only blocks have been accessed more than twice. For instance, in PageRank, almost 100\% of the read-only blocks were accessed more than 20 times. Consequently, the traditional cache replacement mechanism causes read-only data to be fetched to on-chip multiple times, resulting in unnecessary data movement. Based on this phenomenon, this work proposes a corresponding cache management mechanism to reduce the number of \texttt{Read-Only} requests.

\begin{figure}[htbp]
	\vspace{-5pt}
	\centering
	\subfigure[Traditional way for clean victim.]{
		\begin{minipage}[t]{0.45\linewidth}
			\centering
			\includegraphics[width=1.56in]{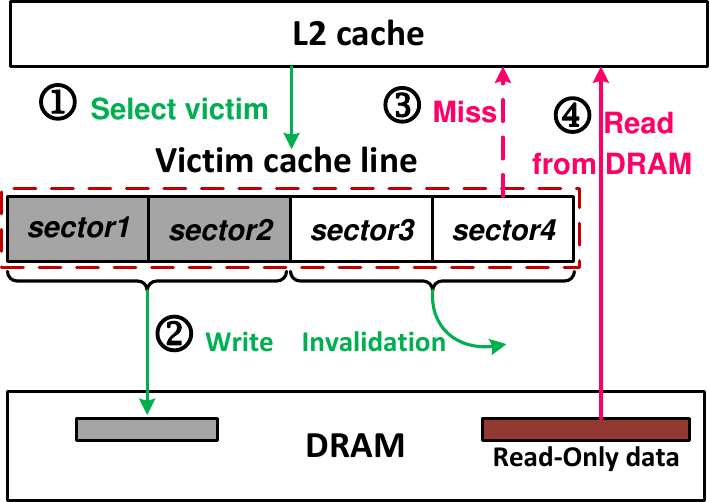}
			%\caption{fig1}
		\end{minipage}%
	}%
	\subfigure[Our FIFO way.]{
		\begin{minipage}[t]{0.55\linewidth}
			\centering
			\includegraphics[width=1.75in]{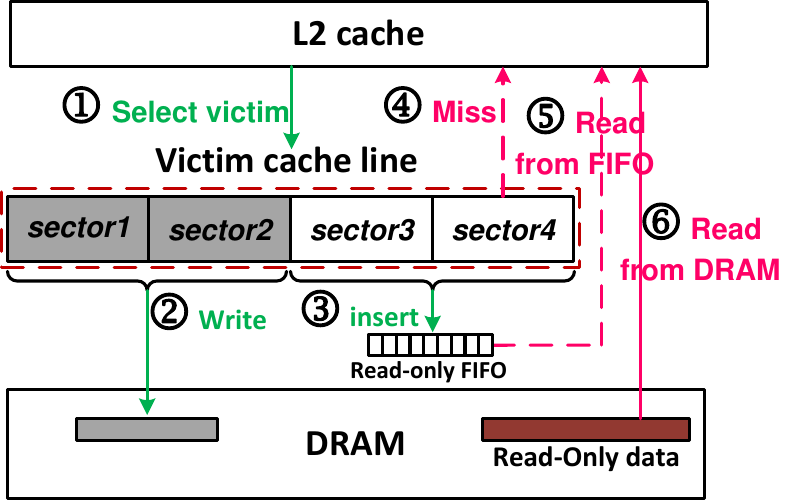}
			%\caption{fig2}
		\end{minipage}%
	}%
	\centering
	\vspace{-5pt}
	\caption{Two methods of processing clean victim sectors.}
	\label{clean_victim}
	\vspace{-5pt}
\end{figure}

For GPU sector caches, when a cache line is selected for eviction, only the modified sectors are written back to DRAM to reduce the bandwidth pressure, while the clean sectors are invalidated. Figure \ref{clean_victim}(a) illustrates the corresponding cache replacement mechanism. Although this approach can reduce the amount of data written to DRAM, it directly invalidates clean read-only data blocks. If these read-only data blocks are needed in the later requests, they must be fetched from DRAM. Clearly, the traditional design is not friendly to workloads like PageRank, BFS, MIS, etc. Therefore, we propose a First In First Out (FIFO) buffer to store these invalidated read-only data, thereby reducing off-chip read requests for these data.

Figure \ref{clean_victim}(b) shows our scheme. When an L2 cache line is selected as a victim \ding{182}, the modified sectors are written back to DRAM \ding{183}, while the clean \& read-only sectors are inserted into the read-only FIFO buffer \ding{184}. When a cache read request comes to the L2 cache, we find if the requested sector is in the L2 cache set or the read-only FIFO buffer. If there is a miss in L2 cache \ding{185} and hit in the FIFO, we read the data in the FIFO \ding{186}. While the sector is not in the FIFO, we must fetch the read-only data from the off-chip DRAM \ding{187}. As for the read-only FIFO, we can use several bloom filters like ~\cite{zhang2019fuse} to make the FIFO buffer approximate full-associate, therefore reducing the hit latency of the FIFO buffer. We assign 16 32B entries for a FIFO, and the FIFO overhead analysis (section \ref{overhead}) and the sensitivity study (section \ref{sensitivity}) are discussed later. In addition, in order to reduce multiple accesses to clean and valid data blocks, the read-only FIFO is not limited to read-only data. If there is clean and valid data in the evicted sector, it can also be placed in the FIFO, thereby reducing the frequency of access from DRAM.

\begin{comment}
	$victim = max(LRU*\alpha + (1-\alpha)*RC_{nor})$
	
	$LRU\in[0, assoc], LRU\in\mathbb{N}$
	
	$RC_{nor} = round((assoc-1)*\dfrac{RC-RC_{min}}{RC_{max}-RC_{min}})$
\end{comment}

\begin{comment}
	\subsection{Metadata Co-location with Compression}
	Due to our sector-granularity memory deduplication, so the metadata overhead is large. We consider to compress the data from 32B to less than 28B to enable store the data in the memory. Therefore, we only execute sector-level deduplication only when the sector is duplicate can can be compressed to less than 28B.
\end{comment}
\subsection{Overhead Analysis}
To implement CMD, we need to add extra hardware and metadata. We classify the overhead into (1) On-chip overhead and (2) Off-chip overhead.

\textbf{\textit{On-chip}}: In order to compute the hash value of a data block, each memory controller needs to integrate one MD5 hardware, which requires 16,259 logic gates ~\cite{lee2010hardware}. For a system with eight memory controllers as shown in Table \ref{system_config}, a total of 130,072 logic gates are required, with an area of 0.014 $mm^2$ at the 16 $nm$ technology node. The MD5 calculation delay for 128-byte data is 228 SM core clock cycles. For the metadata cache, frequently accessed metadata is placed in the memory controller to reduce off-chip access and improve system performance. Specifically, the size of the required hash cache, address cache, type cache, and mask cache per controller is 48KB, 48KB, 5KB, and 10KB, respectively. Therefore, the total on-chip metadata cache size for the GPU system in Table \ref{system_config} is 888KB. Through circuit-level simulator NVsim ~\cite{dong2012nvsim}, it can be simulated that the area of the metadata cache is 0.36 $mm^2$ at the 16 $nm$ technology node. In addition, each L2 cache partition has a read-only FIFO buffer, with the size of 512 bytes. Compared to the 4MB L2 cache, the area overhead is only 0.19\%, and its area is approximately $3.13\times10^{-3} mm^2$. Therefore, the total area overhead of CMD is 0.377 $mm^2$. As a comparison, the area of 1MB cache is 0.41 $mm^2$. In the evaluation part, we compare the scheme with a 5MB L2 cache, which has a close on-chip area with the CMD scheme.

\textbf{\textit{Off-chip}}: In DRAM, each 128-byte block requires a 4-byte size address mapping table. In addition, each block requires a 2-bit type flag and a 4-bit data mask. For read-only sectors, special tags are not needed because the read-only FIFO can handle all clean and valid sectors. The overall capacity overhead for DRAM is 3.71\%.

\label{overhead}
\section{Evaluation}

We implement our scheme in the cycle-accurate GPU simulator GPGPU-Sim 4.0~\cite{bakhoda2009analyzing,khairy2020accel}, and the CUDA version is 11.1. We select several DNN workloads from the Darknet~\cite{darknet13} framework, and the other workloads are from Pannotia~\cite{che2013pannotia}, Rodinia~\cite{che2009rodinia}, ISPASS-2009 ~\cite{bakhoda2009analyzing}. The description of workloads are shown in Table ~\ref{workload}. The detailed GPU system configuration is shown in TABLE \ref{system_config}. As for the energy evaluation, we use the GPU power simulator GPUWattch~\cite{leng2013gpuwattch}. We compare our scheme with the following aggressive techniques:
\begin{itemize}
	\item \textbf{Baseline}: A GPU memory system without optimizations, and the L2 cache size is 4MB.
	\item \textbf{5MB}: A GPU memory system with 5MB L2 cache.
	\item \textbf{BPC}~\cite{kim2016bit}: A state-of-the-art GPU memory compression algorithm. BPC compresses the requested data, reducing the transferred data size between the L2 and DRAM to relieve the bandwidth pressure.
	\item \textbf{BCD}~\cite{park2021bcd}: BCD is a deduplication and compression method designed for CPU memory, which performs differential compression and deduplication on partially duplicated cache lines to reduce the amount of data. We implement BCD on GPU system, and the L2 cache size is 4 MB.
	\item \textbf{ESD}~\cite{du2023esd}: A state-of-the-art secure memory deduplicated method and it adopts the read-verify process for removing the duplicate writes in CPU NVM-based memory. We implement it in the GPU memory system, and the L2 cache size is 4 MB.
	\item \textbf{CMD}: This is our proposed deduplicated GPU memory to avoid duplicate writes, and it integrates the cache-assisted read scheme and the FIFO to reduce the off-chip read requests. The L2 cache size is 4MB, and the on-chip area of CMD is similar to the \texttt{5MB} scheme.
	
\end{itemize}

\begin{table}[htbp]\small
	
	\caption{Application workloads.}
	\vspace{-5pt}
	\begin{center}
		\renewcommand{\arraystretch}{1.0}
		\begin{tabular}{|c|c|c|}
			\hline 
			\bf{Application} & \bf{Dataset} & \bf{Type} \\ 
			\hline 
			darknet & ILSVRC2012 & \multirow{4}*{compute-intensive} \\
			\cline{1-2}
			tiny &  ILSVRC2012 & \\
			\cline{1-2}
			yolo2 & MSCOCO & \\
			\cline{1-2}
			yolo3 & MSCOCO & \\
			\hline
			bfs &ISPASS-2009 & \multirow{9}*{memory-intensive}\\
			\cline{1-2}
			
			mis &Pannotia & \\
			\cline{1-2}
			pagerank &Pannotia & \\
			\cline{1-2}
			color & Pannotia  & \\
			\cline{1-2}
			bp & Rodinia & \\
			\cline{1-2}
			dwt2d & Rodinia & \\
			\cline{1-2}
			kmeans & Rodinia & \\
			\cline{1-2}
			cfd & Rodinia & \\
			\cline{1-2}
			sssp & Pannotia & \\
			\hline
		\end{tabular} 
	\end{center}
	\label{workload} 
	\vspace{-5pt}
\end{table}

\begin{table}[htbp]
	
	\caption{GPU System Configuration.}\small
	\vspace{-5pt}
	\begin{center}
		\renewcommand{\arraystretch}{1.05}
		\begin{tabular}{|c|c|}
			\hline 
			\bf{Component} & \bf{Parameters} \\ 
			\hline 
			Streaming Multiprocessors & 80 SMs, 1365MHz \\
			\hline 
			Warp Size &  32\\
			\hline
			
			Schedulers/Core &  4\\
			
			\hline	
			Number of Threads/Core &  1024 \\
			\hline
			Registers/Core &  65536 \\
			\hline
			Shared Memory/Core & 64KB\\
			\hline
			L1 Data Cache/Core & 64KB, 128B cache line, LRU \\
			\hline
			L2 Data Cache &4MB, 128B cache line, 16-way, LRU \\
			\hline
			Metadata Cache &888KB, 128B cache line, LRU, 20 cycles \\
			\hline
			Interconnection Network & Crossbar, 32B width  \\
			\hline
			Memory Controller &8 MCs, 16 banks/MC, FR-FCFS \\
			\hline
			MD5 hardware &16K gates, 228 cycles/128B \\
			\hline
			\multirow{2}{*}{GDDR6 Timing} & tCL=20, tRP=20, tRC=62, tRAS=50,\\
			& tRCD=20, tRRD=10, tCCD=4, tWR=20 \\
			\hline
		\end{tabular} 
	\end{center}
	\label{system_config} 
	\vspace{-5pt}
\end{table}

\begin{comment}
	\begin{table}[htbp]\footnotesize
		
		\caption{Application benchmarks.}
		\vspace{-10pt}
		\begin{center}
			\renewcommand{\arraystretch}{1.1}
			\begin{tabular}{|c|c|c|}
				\hline 
				\bf{Application} & \bf{Dataset} & \bf{Model/Data Size} \\ 
				\hline 
				Darknet & ILSVRC2012 &28MB \\
				\hline 
				Tiny &  ILSVRC2012 & 4MB\\
				\hline
				
				ResNet18 &  ILSVRC2012& 44MB\\
				
				\hline	
				Extraction &  ILSVRC2012& 90MB \\
				\hline
				Alexnet &  ILSVRC2012 & 238MB\\
				\hline
				Cifarnet & CIFAR10& 30MB\\
				\hline
				YOLO2-Tiny & MSCOCO & 43MB\\
				\hline
				YOLO3-Tiny & MSCOCO & 34MB\\
				\hline
				Breadth First Search &ISPASS-2009 & 61MB\\
				\hline
				Graph Coloring & Pannotia  & 27MB\\
				\hline
				Maximal Independent Set &Pannotia & 44MB\\
				\hline
				Page Rank &Pannotia & 12MB\\
				\hline
			\end{tabular} 
		\end{center}
		\label{workload} 
		\vspace{-10pt}
	\end{table}
\end{comment}
\subsection{Requests Breakdown}

\begin{figure}[htbp]
	\vspace{-5pt}
	\centerline{\includegraphics [width=0.5\textwidth]{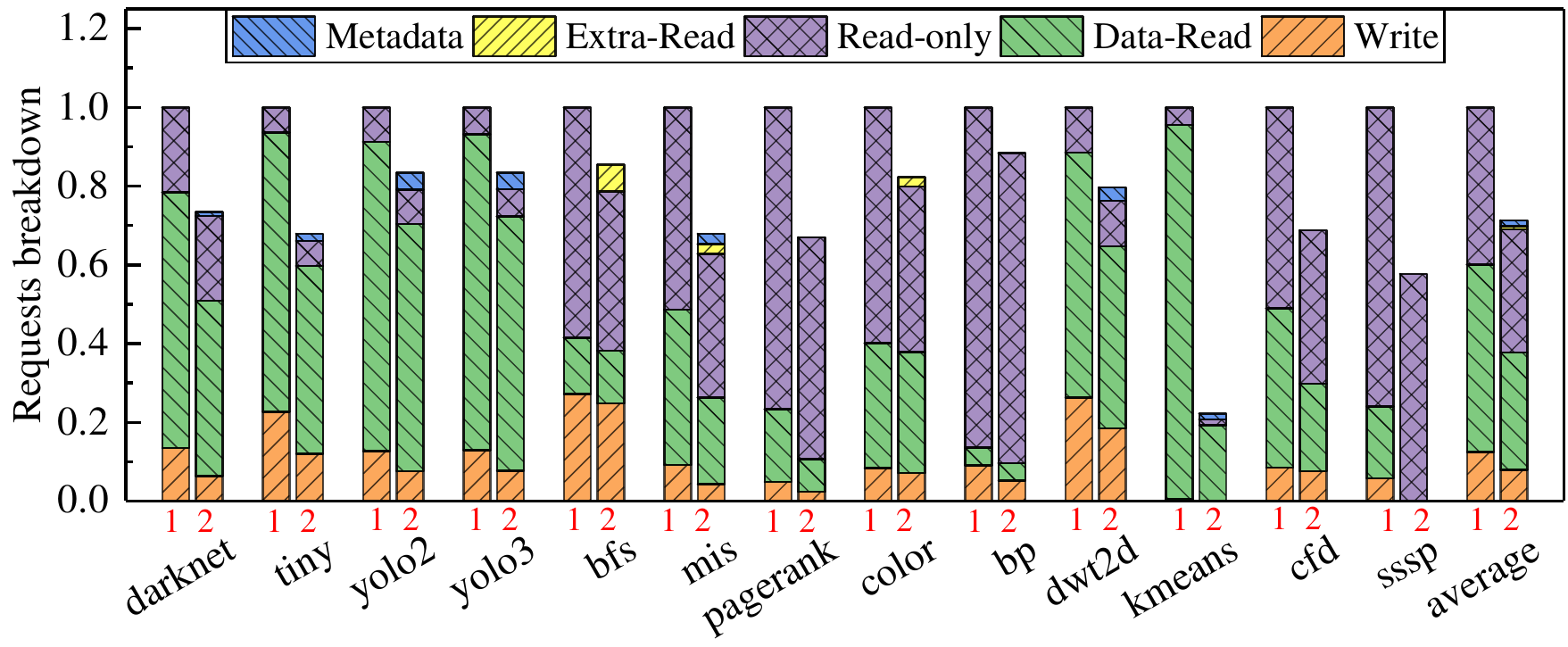}}
	\vspace{-5pt}
	\caption{The requests breakdown of the \textbf{Baseline} and \textbf{CMD}. The left bar \textcolor{red}{1} is \textbf{Baseline}, and the right bar \textcolor{red}{2} is \textbf{CMD}.}
	\label{breakdown}
	\vspace{-5pt}
\end{figure}

Figure \ref{breakdown} shows the request breakdown of the \texttt{Baseline} and our \texttt{CMD} system. Besides the three kinds of requests in Figure \ref{req_ana}, we add the \texttt{Metadata} and \texttt{Dedup-Read} requests. \texttt{Metadata} is the request for querying the metadata from the off-chip DRAM, and \texttt{Dedup-Read} is the extra read request in the write deduplication process. Since \texttt{Dedup-Read} and \texttt{Metadata} requests affect the system performance, but both of them are relatively small compared to the total requests. Compared with the baseline, CMD can reduce the off-chip accesses by 31.01\%. In detail, CMD can reduce \texttt{Write}, \texttt{Data-Read}, and \texttt{Read-Only} request by 35.86\%, 37.60\%, and 21.65\%, respectively.
\subsection{Performance}

\begin{figure}[htbp]
	\vspace{-5pt}
	\centerline{\includegraphics [width=0.5\textwidth]{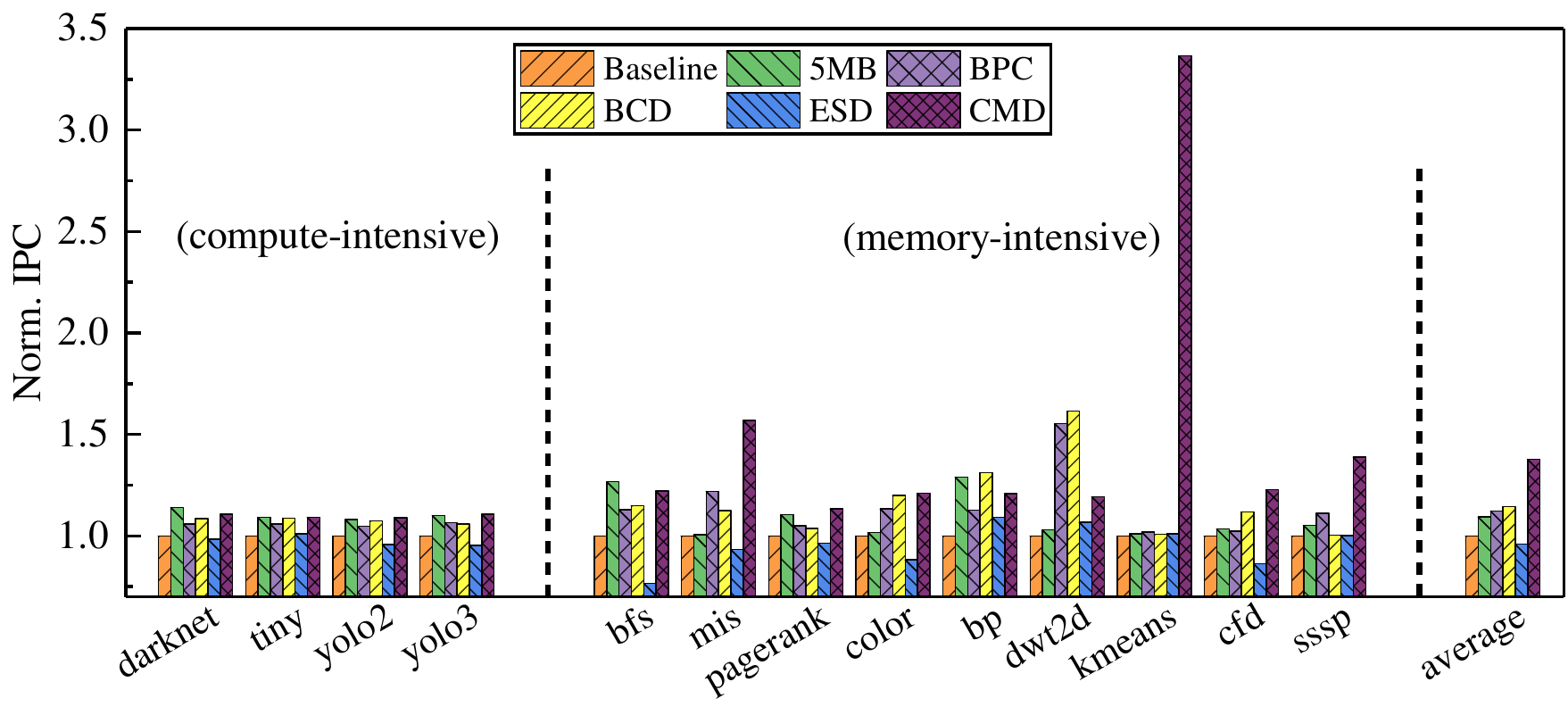}}
	\vspace{-5pt}
	\caption{The normalized performance of several schemes.}
	\label{ipc}
	\vspace{-5pt}
\end{figure}

We use Instruction Per Cycle (IPC) to characterize the performance of the GPU system. Figure ~\ref{ipc} shows the normalized performance of different schemes. Compared to the \texttt{Baseline}, the \texttt{5MB}, \texttt{BPC}, \texttt{BCD}, \texttt{ESD}, and \texttt{CMD} can improve performance by 9.42\%, 12.30\%, 14.38\%, -3.98\%, and 37.79\%, respectively. For \texttt{5MB}, a 5 MB L2 cache is used, reducing the L2 cache miss rate and thus reducing off-chip access to improve overall system performance. For \texttt{BPC} and \texttt{BCD}, they can significantly reduce the data transferred between the L2 cache and DRAM, leading to performance improvement. \texttt{ESD} can reduce redundant data writes, but the read\&verify process increases the bandwidth pressure, leading to performance degradation. For \texttt{CMD}, as shown in the test results in Figure ~\ref{breakdown}, it can significantly reduce the off-chip accesses and thus improve system performance. In addition, compared to the \texttt{5MB} scheme, \texttt{CMD} has a close on-chip area, and \texttt{CMD} can improve performance by 25.92\% compared to the \texttt{5MB} scheme. Besides, as for the memory-intensive workloads, \texttt{5MB}, \texttt{BPC}, \texttt{BCD}, \texttt{ESD}, and \texttt{CMD} can improve performance by 9.02\%, 15.19\%, 17.37\%, -4.66\%, and 50.18\%, respectively. As for the compute-intensive workloads, \texttt{CMD} can obtain an average performance improvement of 9.91\%, which is similar to the \texttt{5MB} scheme.

\begin{figure}[htbp]
	\vspace{-5pt}
	\centerline{\includegraphics [width=0.5\textwidth]{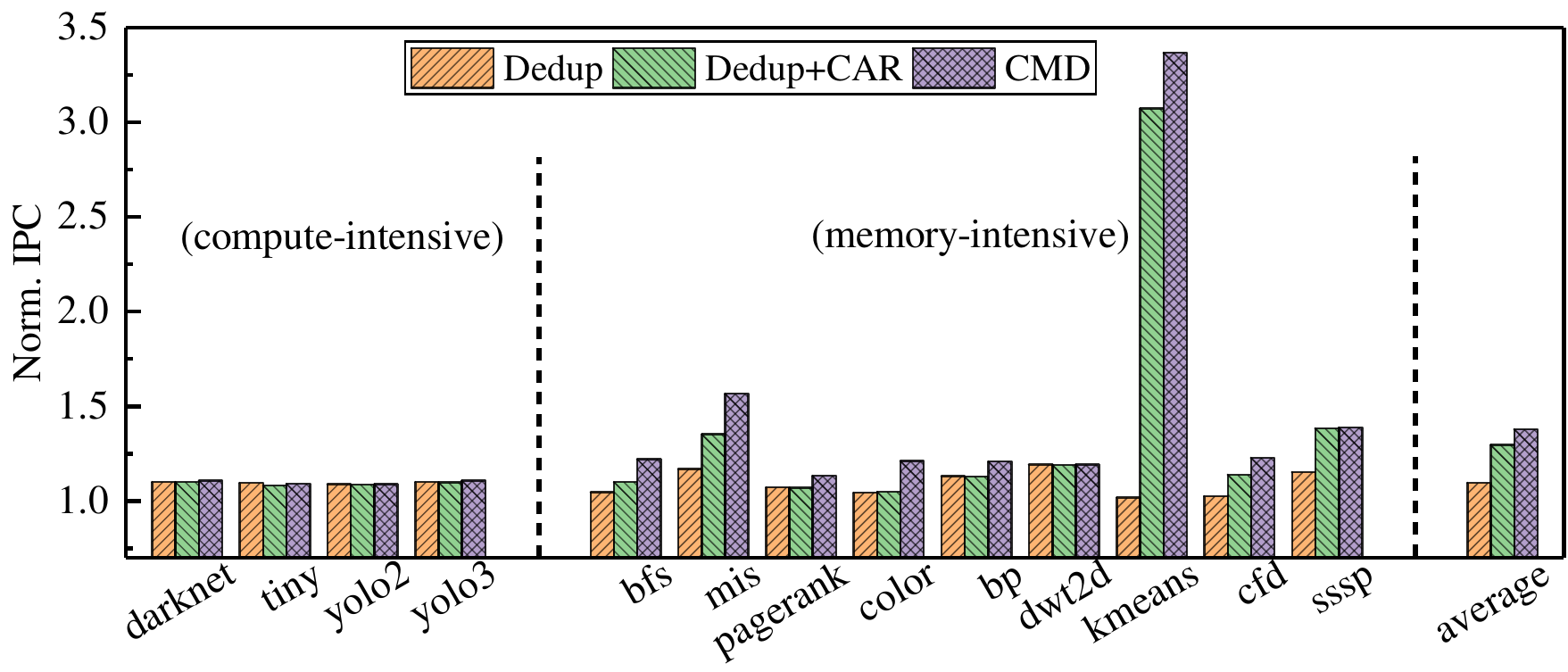}}
	\vspace{-5pt}
	\caption{The normalized performance of several schemes.}
	\label{ipc_breakdown}
	\vspace{-5pt}
\end{figure}

To better analyze the performance gains of various optimization techniques in CMD, Figure \ref{ipc_breakdown} shows the normalized performance gains of the three techniques in CMD. For the \texttt{Dedup} method, which primarily reduces the off-chip duplicate data writes, it can improve the IPC by 9.52\%. After integrating the Cache Assisted Read (CAR) method, the system performance can be improved by 29.62\%. This is because CAR can significantly reduce the off-chip duplicate \texttt{Data-Read} requests. For some workloads, such as tiny and yolo2, the performance slightly declines after integrating the CAR method since the cache-assisted read approach needs to detect whether the reference block data is in the L2 cache, and this process requires access to the L2 cache, introducing additional latency. Ultimately, incorporating the read-only FIFO buffering strategy further improves performance. For workloads like darknet, tiny, yolo2, and dwt2d, since the hit rate for read-only data blocks in the read-only FIFO is low, thus there is no apparent performance benefit. As for the memory-intensive workloads, \texttt{Dedup}, \texttt{Dedup+CAR}, and \texttt{CMD} can improve the performance by 9.46\%, 38.71\%, and 50.18\% on average. While for compute-intensive workloads, the IPC improvements are not significant. 

\subsection{Energy Consumption}
Our work primarily focuses on the optimization of L2 cache and off-chip DRAM. Therefore, we only consider the energy of the L2 cache, memory controller, and DRAM. In addition, the energy overhead of the metadata cache and MD5 hardware are accounted for in the memory controller. To obtain energy consumption data, we employ the GPU power simulator GPUWattch to obtain the power consumption of the L2 cache, memory controller, and DRAM. Subsequently, energy consumption can be calculated as the product of power and execution time.

\begin{figure}[htbp]
	\vspace{-5pt}
	\centerline{\includegraphics [width=0.5\textwidth]{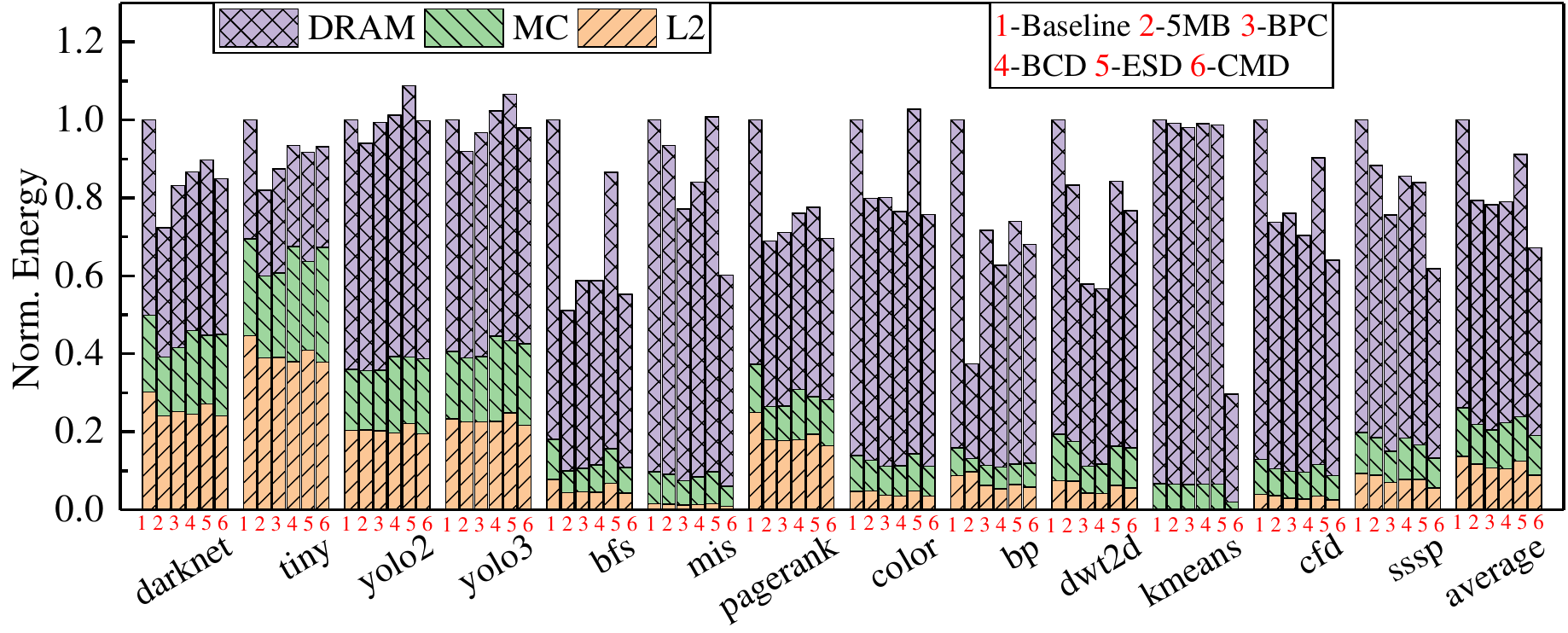}}
	\vspace{-5pt}
	\caption{The energy consumption of DRAM, MC and L2 cache.}
	\label{energy}
	\vspace{-5pt}
\end{figure}

Figure \ref{energy} presents the normalized energy consumption test results. Compared to the \texttt{Baseline}, \texttt{5MB}, \texttt{BPC}, \texttt{BCD}, \texttt{ESD}, and \texttt{CMD} can reduce energy consumption by 20.69\%, 21.78\%, 21.02\%, 8.80\%, and 32.78\%, respectively. For \texttt{5MB} scheme, the larger L2 cache can reduce off-chip DRAM energy consumption. For the \texttt{BPC} and \texttt{BCD} schemes, they can reduce the amount of data written to DRAM, resulting in less energy consumption. For the \texttt{ESD} scheme, it can reduce the duplicate writes to DRAM with the read\&verify process. For the \texttt{CMD} scheme, it can significantly reduce the power of DRAM and L2 cache. Besides, CMD decreases the program's execution time as well, thus reducing the runtime energy consumption.
\begin{figure*}[htbp]
	\vspace{-5pt}
	\centering
	\subfigure[Deduplication ratio with different size.]{
		\begin{minipage}[t]{0.25\linewidth}
			\centering
			\includegraphics[width=1.75in]{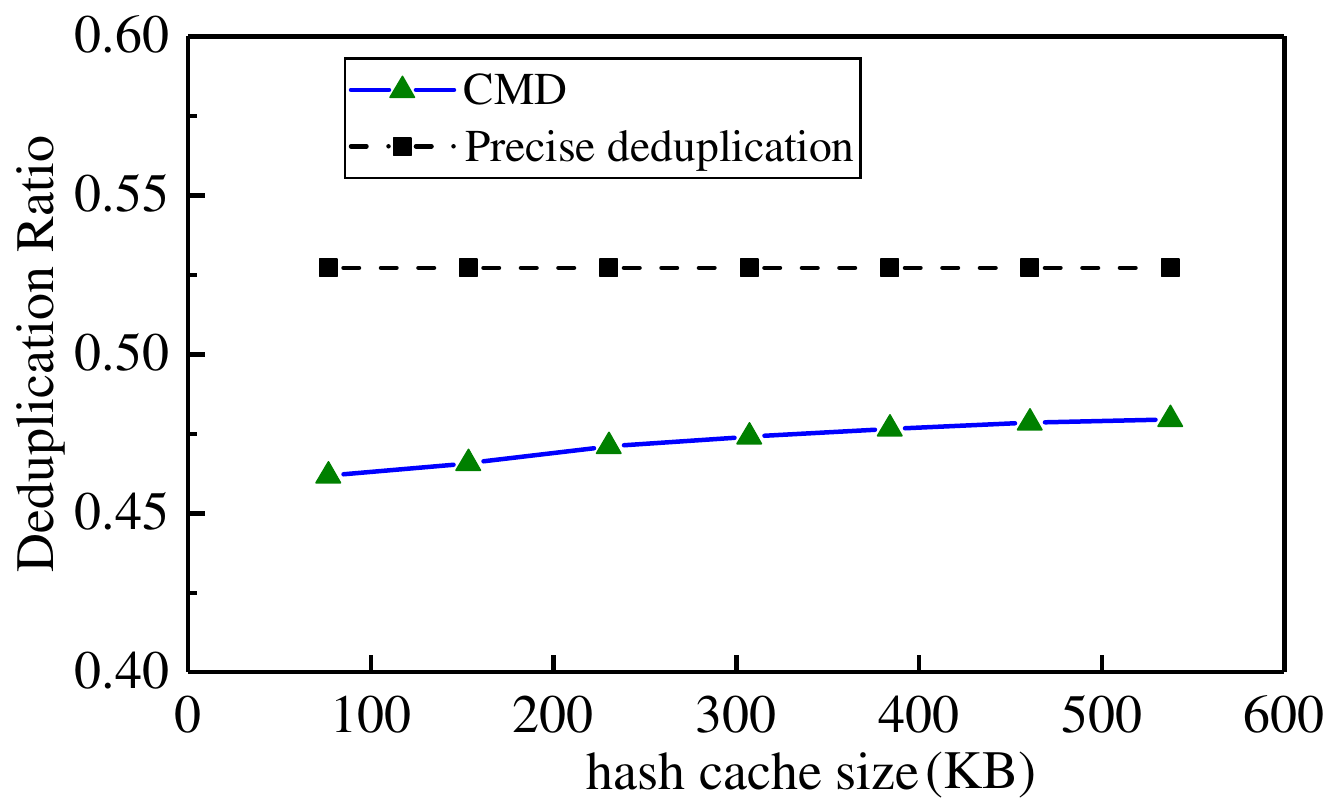}
			%\caption{fig1}
		\end{minipage}%
	}%
	\subfigure[Hit rate of address cache.]{
		\begin{minipage}[t]{0.25\linewidth}
			\centering
			\includegraphics[width=1.75in]{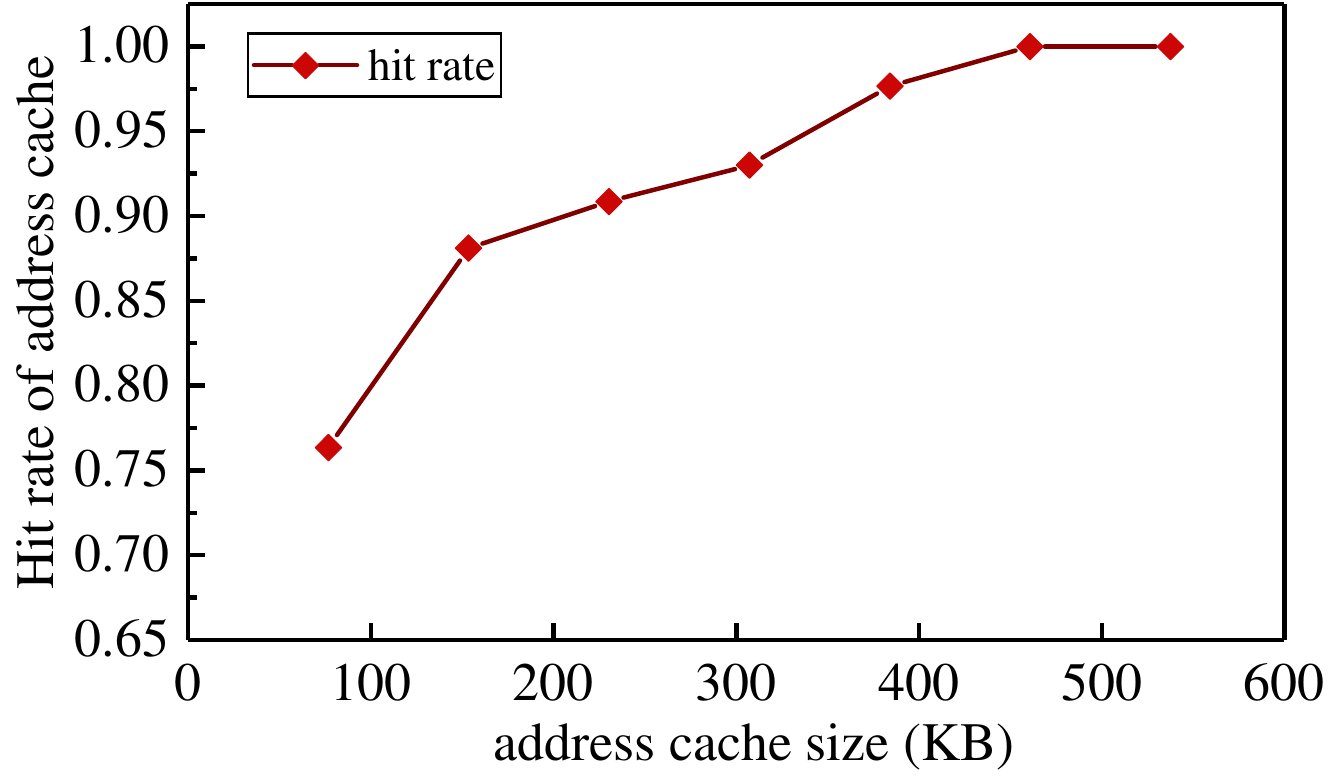}
			%\caption{fig2}
		\end{minipage}%
	}%
	\subfigure[Hit rate of mask cache.]{
		\begin{minipage}[t]{0.25\linewidth}
			\centering
			\includegraphics[width=1.75in]{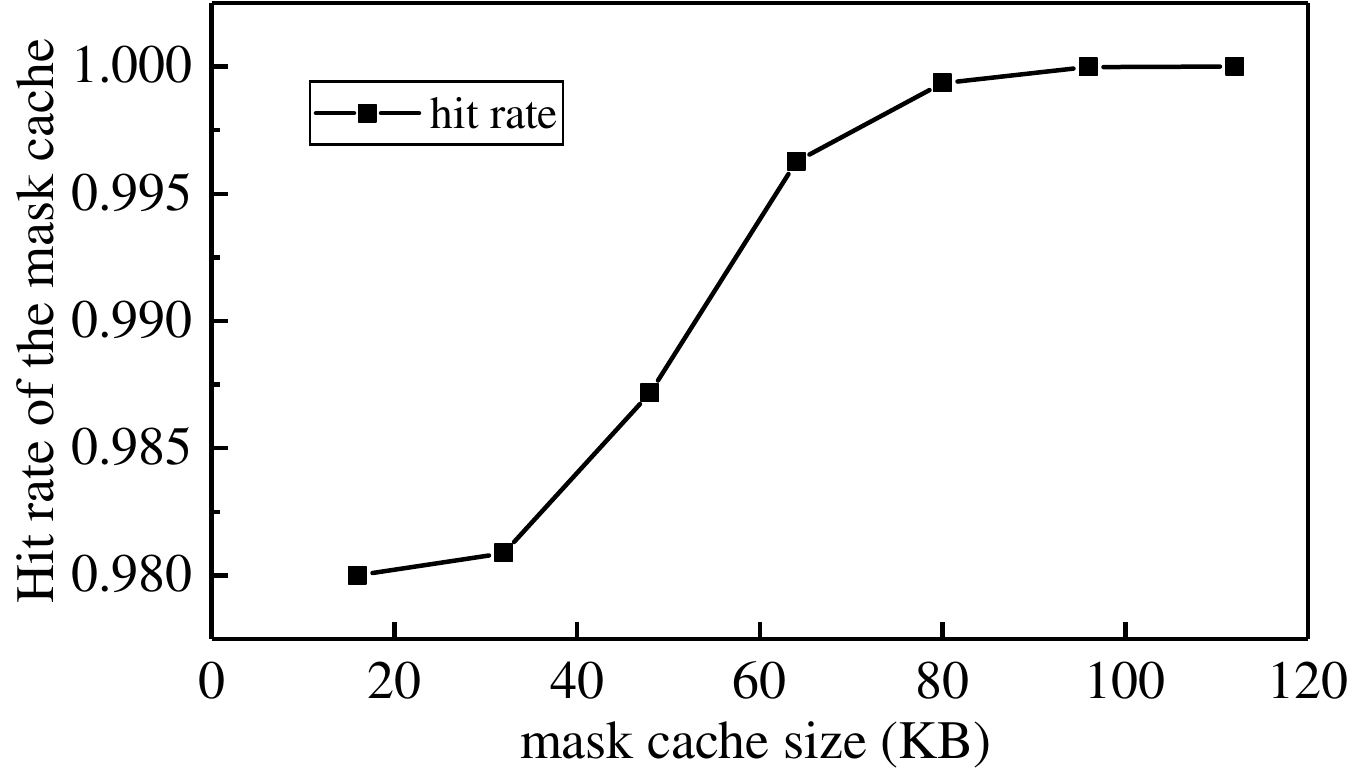}
			%\caption{fig2}
		\end{minipage}%
	}%
	\subfigure[Hit rate of type cache.]{
		\begin{minipage}[t]{0.25\linewidth}
			\centering
			\includegraphics[width=1.75in]{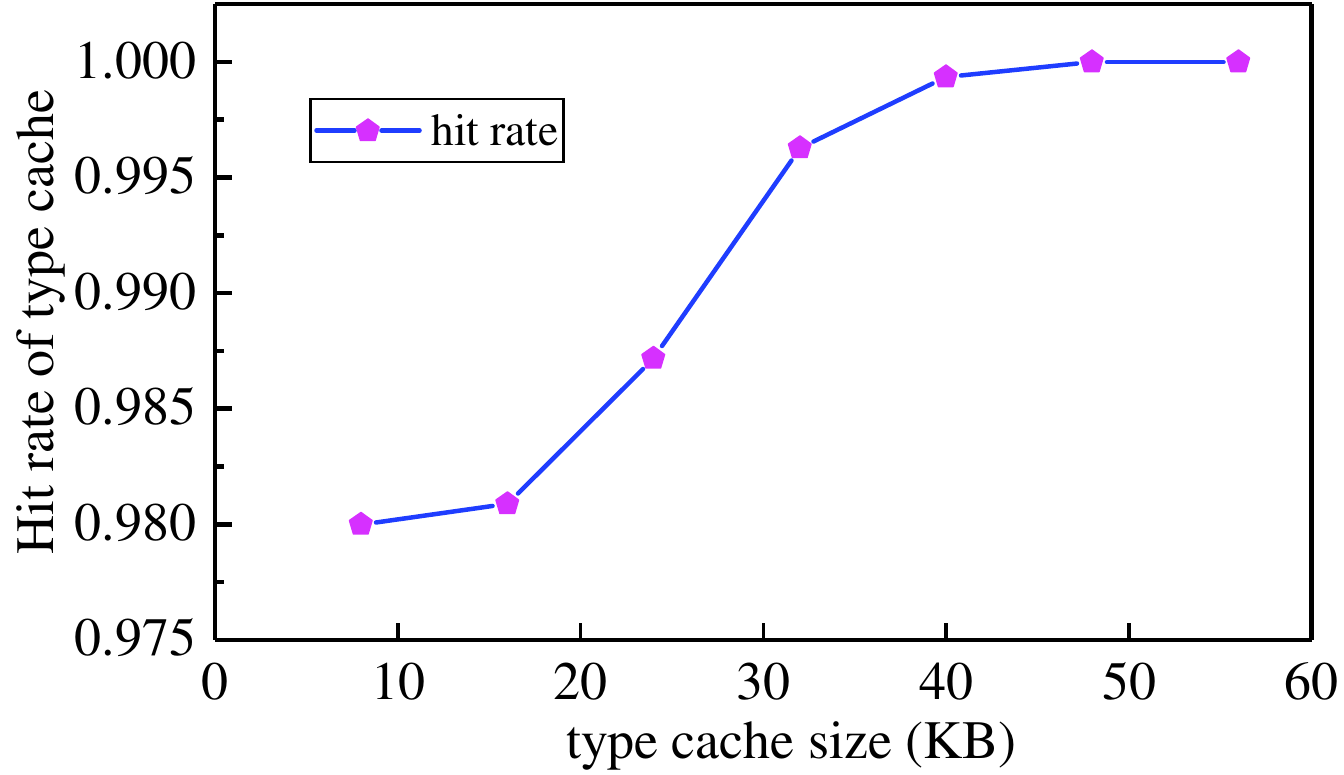}
			%\caption{fig2}
		\end{minipage}%
	}%
	\centering
	\vspace{-10pt}
	\caption{The sensitivity study of the metadata cache size.}
	\vspace{-5pt}
	\label{metadata_sensitivity}
\end{figure*}
\subsection{Sensitivity Study}

In the CMD design, there are several key parameters that will directly affect the overall benefit. Therefore, we conduct the design parameter sensitivity study.

\textbf{(1) Metadata cache size sensitivity study.} Figure \ref{metadata_sensitivity}(a) illustrates the relationship between the deduplication ratio and the hash cache size. When the cache capacity is 77 KB, 46.17\% of duplicate writes can be reduced. When the cache capacity is increased to 538 KB, 47.95\% of duplicate writes can be reduced. The deduplication ratio is not sensitive to cache capacity, and this is because a large number of duplicate blocks are mainly mapped to a small portion of reference blocks. In addition, the CMD scheme adopts partial deduplication. Compared to the exact deduplication, CMD can achieve 90\% of its deduplication effect, but exact deduplication involves a large amount of off-chip hash data access, which severely degrades performance.

Figure \ref{metadata_sensitivity}(b) displays the relationship between the address cache hit rate and cache size. When the cache capacity is 384 KB, the cache hit rate can reach 97.66\%. Figure \ref{metadata_sensitivity}(c) shows the relationship between the mask cache hit rate and capacity. Since the mask data is small, with only 4 bits per data block, a small capacity cache can achieve a high hit rate. When the cache capacity is 80 KB, the hit rate is 99.93\%. Besides the large mask cache, the cache block prefetch granularity is 32B, which includes 64 mask data, resulting in a cache hit rate close to 100\%. Figure \ref{metadata_sensitivity}(d) indicates the relationship between the type cache hit rate and capacity. Similar to the mask cache, since each memory block's type cache data is only 2 bits, a small capacity can achieve a high cache hit rate. In the CMD scheme, to balance the performance and capacity overhead of metadata cache, the total capacities of hash cache, address cache, mask cache, and type cache are set to 384 KB, 384 KB, 80 KB, and 40 KB, respectively. The total metadata cache is 888 KB.

\begin{figure}[htbp]
	\vspace{-15pt}
	\centerline{\includegraphics [width=0.5\textwidth]{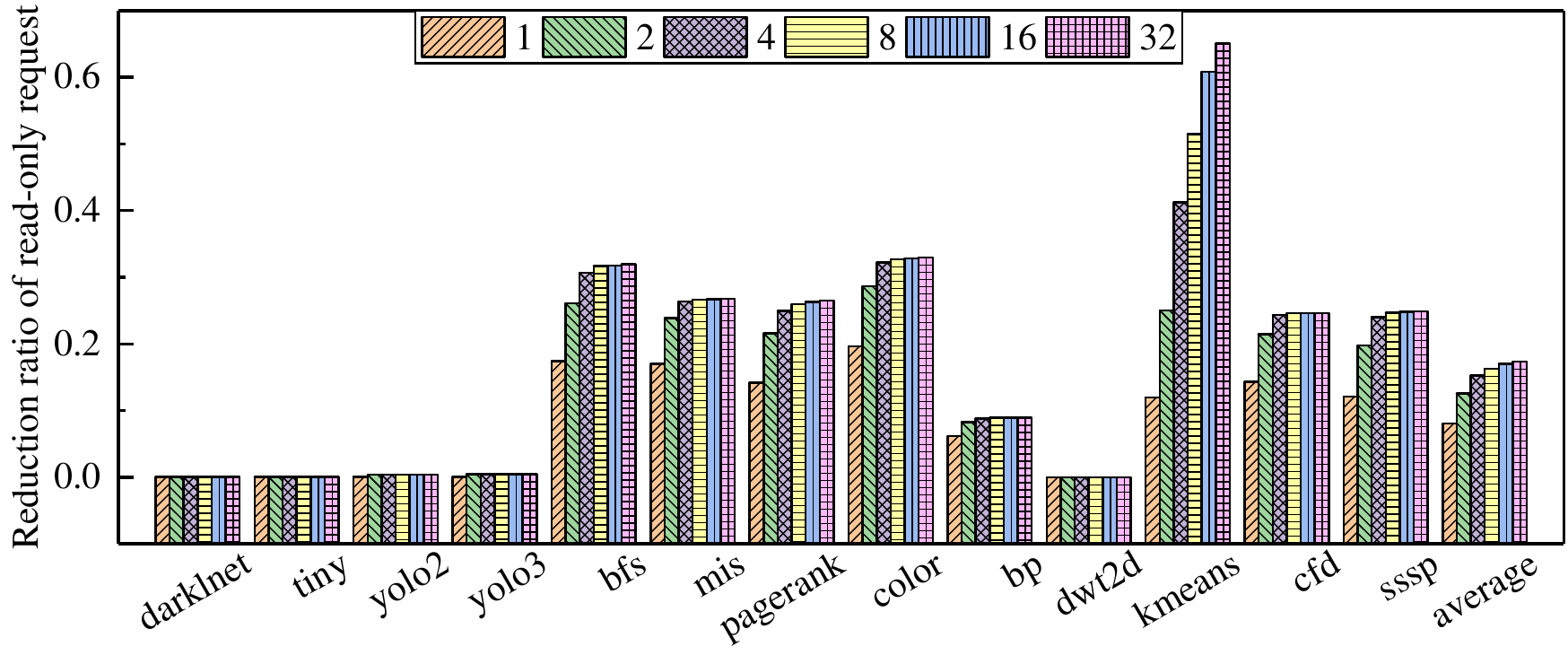}}
	\vspace{-5pt}
	\caption{The sensitivity study of FIFO size.}
	\label{fifo_sensitivity}
	\vspace{-5pt}
\end{figure}

\textbf{(2) Read-only FIFO size sensitivity study.} In optimizing read-only data requests, the size of the FIFO buffer affects the overall effectiveness of the scheme. Figure \ref{fifo_sensitivity} presents the test results of the relationship between FIFO size and read-only request reduction ratio. For some workloads such as darknet, tiny, yolo2, yolo3, and dwt2d, the read-only FIFO cannot reduce read-only requests. This is because the read-only requests are mostly used only once or twice. Consequently, there are few repeated reads from DRAM, so the FIFO buffer cannot reduce read-only requests for these applications.

However, for some workloads like bfs, mis, pagerank, and color, their read-only data is accessed multiple times. Therefore, the FIFO can buffer a large number of read-only data blocks, thus reducing off-chip DRAM access for these data blocks. On average, when the FIFO size is set to 1, 2, 4, 8, 16, and 32, the read-only requests can be reduced by 8.05\%, 12.56\%, 15.27\%, 16.27\%, 17.00\%, and 17.33\%, respectively. To ensure the effectiveness of the FIFO scheme while keeping low hardware overhead, this work sets the FIFO size to 16.

\label{sensitivity}
\subsection{Discussion with Scalability}
The CMD architecture aims to reduce duplicate data read and write operations and read-only requests, thereby optimizing GPU memory bandwidth pressure. Compression schemes are frequently employed to alleviate GPU memory bandwidth pressure and improve performance. Therefore, the CMD scheme can combine with compression methods to enhance performance. When integrated with the BPC scheme, if the data written to DRAM is non-duplicate, BPC can be used to reduce the data size.

\begin{figure}[htbp]
	\vspace{-10pt}
	\centerline{\includegraphics [width=0.5\textwidth]{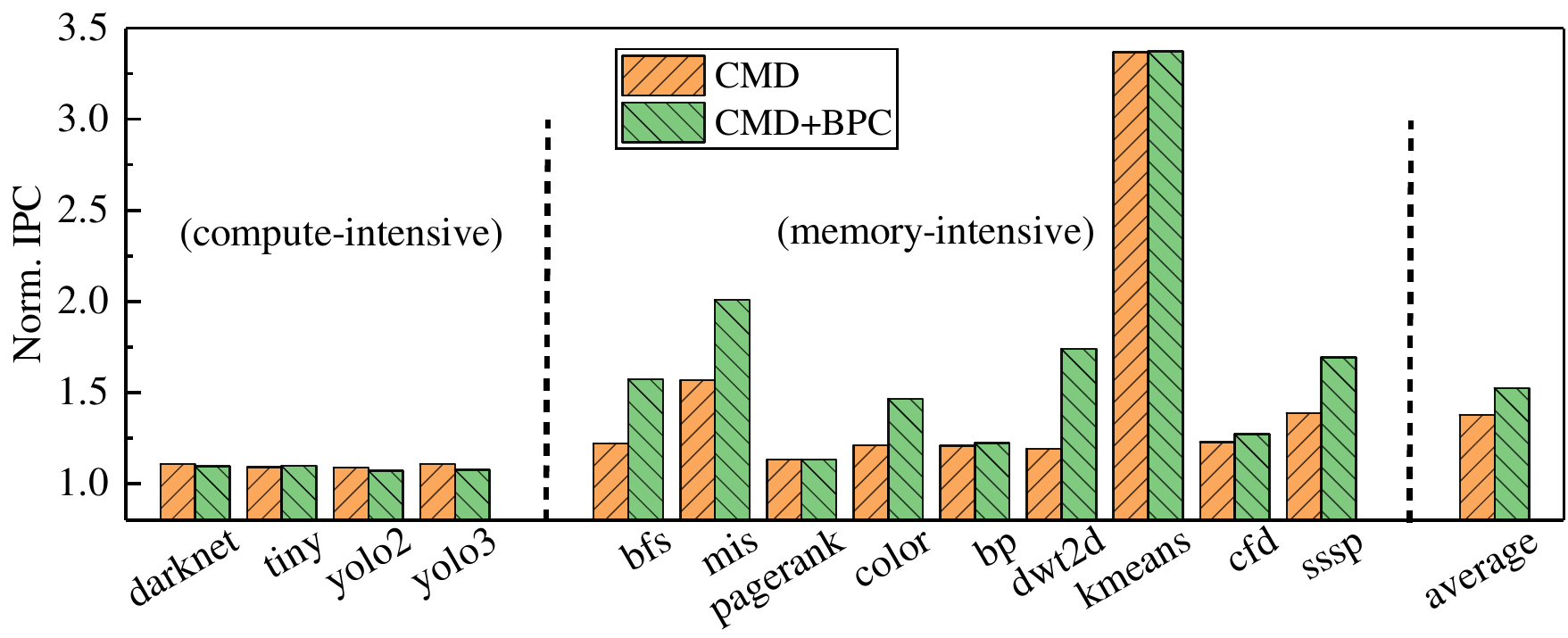}}
	\vspace{-5pt}
	\caption{The normalized IPC combined with BPC.}
	\label{cmd_bpc}
	\vspace{-15pt}
\end{figure}

Figure \ref{cmd_bpc} displays the normalized performance of the CMD combined with BPC, while the results are normalized to the \texttt{Baseline} scheme. The results show that CMD+BPC can achieve a 52.53\% performance improvement. For the compute-intensive application workloads, such as darknet, tiny, etc., there is no significant improvement in IPC performance, as the performance bottleneck is not in the bandwidth but in the GPU compute cores. However, for memory-sensitive application workloads like bfs, mis, and color, CMD+BPC can improve the performance by 72.05\% on average. 
\begin{comment}
	\begin{figure}[htbp]
		
		\centerline{\includegraphics [width=0.5\textwidth]{./plot/cmd_bpc.pdf}}
		\vspace{-5pt}
		\caption{The performance while integrating BPC into CMD.}
		\label{cmd_bpc_ipc}
		
	\end{figure}
\end{comment}

\section{Related Work}
\textbf{GPU Memory Compression and Performance Optimization.} Memory compression techniques are usually used to relieve the GPU bandwidth pressure. Sathish et al. ~\cite{sathish2012lossless} proposed a lossy and lossless memory compression architecture to reduce the transferred data size. As for the lossy compression technique, it truncated the insignificant decimal bits in floating-point numbers. Kim et al. ~\cite{kim2016bit} proposed Bit Plane Compression (BPC) algorithm. BPC performed bit-level stratification on the data, and XOR operation was used to convert most of the data to bit "0", thereby significantly reducing the information entropy and improving the data compression ratio. Sullivan et al. proposed a Buddy compression~\cite{choukse2020buddy} method to expand the effective GPU memory capacity. Darabi et al.~\cite{darabi2022morpheus} proposes to allocate on-chip cache resources from idle cores to the GPU's last-level cache, effectively expanding the capacity of the last-level cache to improve system performance. As for the GPU memory compression schemes, they only consider the intra-line information, which cannot remove the duplicate blocks. As for the cache-level optimizations, it's orthogonal to our work.

\textbf{CPU Memory Deduplication.} Currently, some CPU cache line-level memory deduplication techniques are proposed. DeWrite~\cite{zuo2018improving} proposed to use the weak hash and read\&verify process to remove the duplicate writes to NVM main memory. ESD~\cite{du2023esd} proposed to use the Error Correct Code (ECC) of the last level cache as the fingerprint. Besides, ESD adopted a selective deduplication mechanism to avoid large off-chip accesses. BCD~\cite{park2021bcd} proposed to perform differential compression and deduplication on partially duplicated cache lines to reduce the amount of data. CMD is the first work that proposes the GPU memory deduplication architecture.
\section{Conclusion}

Massive off-chip accesses in GPUs limit the system's performance. By analyzing the off-chip accesses, we classify the off-chip requests into three types: (1) Write, (2) Data-Read, and (3) Read-Only. In this work, we propose a cache-assisted GPU memory deduplication architecture named \textbf{CMD} to reduce the three types of off-chip requests. CMD includes three key design contributions: (1) A novel GPU memory deduplication architecture that reduces the inter-duplicate and intra-duplicate write requests. (2) Besides the duplicate writes, we propose a cache-assisted read scheme to reduce the reads to duplicate data with the assistance of metadata cache and L2 cache. (3) For the read-only data, most clean sector victims are re-referenced more than twice. Therefore, we add a full-associate FIFO to accommodate the read-only victims to reduce the re-reference counts. Experiments show that CMD can decrease the off-chip access by 31.01\%, reduce the energy by 32.78\% and improve performance by 37.79\%. Besides, CMD can improve the performance of memory-intensive workloads by 50.18\%.
\normalem
\bibliographystyle{IEEEtran}

\bibliography{references}

\begin{thebibliography}{10}
\providecommand{\url}[1]{#1}
\csname url@samestyle\endcsname
\providecommand{\newblock}{\relax}
\providecommand{\bibinfo}[2]{#2}
\providecommand{\BIBentrySTDinterwordspacing}{\spaceskip=0pt\relax}
\providecommand{\BIBentryALTinterwordstretchfactor}{4}
\providecommand{\BIBentryALTinterwordspacing}{\spaceskip=\fontdimen2\font plus
\BIBentryALTinterwordstretchfactor\fontdimen3\font minus
  \fontdimen4\font\relax}
\providecommand{\BIBforeignlanguage}[2]{{%
\expandafter\ifx\csname l@#1\endcsname\relax
\typeout{** WARNING: IEEEtran.bst: No hyphenation pattern has been}%
\typeout{** loaded for the language `#1'. Using the pattern for}%
\typeout{** the default language instead.}%
\else
\language=\csname l@#1\endcsname
\fi
#2}}
\providecommand{\BIBdecl}{\relax}
\BIBdecl

\bibitem{zhao2019adaptive}
X.~Zhao, A.~Adileh, Z.~Yu, Z.~Wang, A.~Jaleel, and L.~Eeckhout, ``Adaptive
  memory-side last-level gpu caching,'' in \emph{Proceedings of the 46th
  International Symposium on Computer Architecture}, 2019, pp. 411--423.

\bibitem{zhao2022low}
W.~Zhao, J.~Xu, X.~Wei, B.~Wu, C.~Wang, W.~Zhu, W.~Tong, D.~Feng, and J.~Liu,
  ``A low latency and high endurance mlc stt-mram based cache system,''
  \emph{IEEE Transactions on Computer-Aided Design of Integrated Circuits and
  Systems}, 2022.

\bibitem{zhang2019fuse}
J.~Zhang, M.~Jung, and M.~Kandemir, ``Fuse: Fusing stt-mram into gpus to
  alleviate off-chip memory access overheads,'' in \emph{2019 IEEE
  International Symposium on High Performance Computer Architecture
  (HPCA)}.\hskip 1em plus 0.5em minus 0.4em\relax IEEE, 2019, pp. 426--439.

\bibitem{inci2021deepnvm++}
A.~Inci, M.~M. Isgenc, and D.~Marculescu, ``Deepnvm++: Cross-layer modeling and
  optimization framework of non-volatile memories for deep learning,''
  \emph{IEEE Transactions on Computer-Aided Design of Integrated Circuits and
  Systems}, 2021.

\bibitem{darabi2022morpheus}
S.~Darabi, M.~Sadrosadati, N.~Akbarzadeh, J.~Lindegger, M.~Hosseini, J.~Park,
  J.~G{\'o}mez-Luna, O.~Mutlu, and H.~Sarbazi-Azad, ``Morpheus: Extending the
  last level cache capacity in gpu systems using idle gpu core resources,'' in
  \emph{2022 55th IEEE/ACM International Symposium on Microarchitecture
  (MICRO)}.\hskip 1em plus 0.5em minus 0.4em\relax IEEE, 2022, pp. 228--244.

\bibitem{sathish2012lossless}
V.~Sathish, M.~J. Schulte, and N.~S. Kim, ``Lossless and lossy memory i/o link
  compression for improving performance of gpgpu workloads,'' in
  \emph{Proceedings of the 21st international conference on Parallel
  architectures and compilation techniques}, 2012, pp. 325--334.

\bibitem{kim2016bit}
J.~Kim, M.~Sullivan, E.~Choukse, and M.~Erez, ``Bit-plane compression:
  Transforming data for better compression in many-core architectures,'' in
  \emph{2016 ACM/IEEE 43rd Annual International Symposium on Computer
  Architecture (ISCA)}.\hskip 1em plus 0.5em minus 0.4em\relax IEEE, 2016, pp.
  329--340.

\bibitem{rhu2018compressing}
M.~Rhu, M.~O'Connor, N.~Chatterjee, J.~Pool, Y.~Kwon, and S.~W. Keckler,
  ``Compressing dma engine: Leveraging activation sparsity for training deep
  neural networks,'' in \emph{2018 IEEE International Symposium on High
  Performance Computer Architecture (HPCA)}.\hskip 1em plus 0.5em minus
  0.4em\relax IEEE, 2018, pp. 78--91.

\bibitem{choukse2020buddy}
E.~Choukse, M.~B. Sullivan, M.~O’Connor, M.~Erez, J.~Pool, D.~Nellans, and
  S.~W. Keckler, ``Buddy compression: Enabling larger memory for deep learning
  and hpc workloads on gpus,'' in \emph{2020 ACM/IEEE 47th Annual International
  Symposium on Computer Architecture (ISCA)}.\hskip 1em plus 0.5em minus
  0.4em\relax IEEE, 2020, pp. 926--939.

\bibitem{zuo2018improving}
P.~Zuo, Y.~Hua, M.~Zhao, W.~Zhou, and Y.~Guo, ``Improving the performance and
  endurance of encrypted non-volatile main memory through deduplicating
  writes,'' in \emph{2018 51st Annual IEEE/ACM International Symposium on
  Microarchitecture (MICRO)}.\hskip 1em plus 0.5em minus 0.4em\relax IEEE,
  2018, pp. 442--454.

\bibitem{park2021bcd}
S.~Park, I.~Kang, Y.~Moon, J.~H. Ahn, and G.~E. Suh, ``Bcd deduplication:
  Effective memory compression using partial cache-line deduplication,'' in
  \emph{Proceedings of the 26th ACM International Conference on Architectural
  Support for Programming Languages and Operating Systems}, 2021, pp. 52--64.

\bibitem{bakhoda2009analyzing}
A.~Bakhoda, G.~L. Yuan, W.~W. Fung, H.~Wong, and T.~M. Aamodt, ``Analyzing cuda
  workloads using a detailed gpu simulator,'' in \emph{2009 IEEE international
  symposium on performance analysis of systems and software}.\hskip 1em plus
  0.5em minus 0.4em\relax IEEE, 2009, pp. 163--174.

\bibitem{khairy2020accel}
M.~Khairy, Z.~Shen, T.~M. Aamodt, and T.~G. Rogers, ``Accel-sim: An extensible
  simulation framework for validated gpu modeling,'' in \emph{2020 ACM/IEEE
  47th Annual International Symposium on Computer Architecture (ISCA)}.\hskip
  1em plus 0.5em minus 0.4em\relax IEEE, 2020, pp. 473--486.

\bibitem{sector_cache}
``Kernel profiling guide,''
  \url{https://docs.nvidia.com/nsight-compute/pdf/ProfilingGuide.pdf}, 2022.

\bibitem{xia2016comprehensive}
W.~Xia, H.~Jiang, D.~Feng, F.~Douglis, P.~Shilane, Y.~Hua, M.~Fu, Y.~Zhang, and
  Y.~Zhou, ``A comprehensive study of the past, present, and future of data
  deduplication,'' \emph{Proceedings of the IEEE}, vol. 104, no.~9, pp.
  1681--1710, 2016.

\bibitem{rodeh2003zfs}
O.~Rodeh and A.~Teperman, ``zfs-a scalable distributed file system using object
  disks,'' in \emph{20th IEEE/11th NASA Goddard Conference on Mass Storage
  Systems and Technologies, 2003.(MSST 2003). Proceedings.}\hskip 1em plus
  0.5em minus 0.4em\relax IEEE, 2003, pp. 207--218.

\bibitem{drago2012inside}
I.~Drago, M.~Mellia, M.~M.~Munafo, A.~Sperotto, R.~Sadre, and A.~Pras, ``Inside
  dropbox: understanding personal cloud storage services,'' in
  \emph{Proceedings of the 2012 internet measurement conference}, 2012, pp.
  481--494.

\bibitem{rivest1992md5}
R.~Rivest, ``The md5 message-digest algorithm,'' Tech. Rep., 1992.

\bibitem{jarvinen2005hardware}
K.~Jarvinen, M.~Tommiska, and J.~Skytta, ``Hardware implementation analysis of
  the md5 hash algorithm,'' in \emph{Proceedings of the 38th annual Hawaii
  international conference on system sciences}.\hskip 1em plus 0.5em minus
  0.4em\relax IEEE, 2005, pp. 298a--298a.

\bibitem{huo2015high}
Y.~Huo, X.~Li, W.~Wang, and D.~Liu, ``High performance table-based architecture
  for parallel crc calculation,'' in \emph{The 21st IEEE International Workshop
  on Local and Metropolitan Area Networks}.\hskip 1em plus 0.5em minus
  0.4em\relax IEEE, 2015, pp. 1--6.

\bibitem{du2023esd}
C.~Du, S.~Wu, J.~Wu, B.~Mao, and S.~Wang, ``Esd: An ecc-assisted and selective
  deduplication for encrypted non-volatile main memory,'' in \emph{2023 IEEE
  International Symposium on High-Performance Computer Architecture
  (HPCA)}.\hskip 1em plus 0.5em minus 0.4em\relax IEEE, 2023, pp. 977--990.

\bibitem{lee2010hardware}
Y.~K. Lee, M.~Kne{\v{z}}evi{\'c}, and I.~M. Verbauwhede, ``Hardware design for
  hash functions,'' \emph{Secure Integrated Circuits and Systems}, pp. 79--104,
  2010.

\bibitem{dong2012nvsim}
X.~Dong, C.~Xu, Y.~Xie, and N.~P. Jouppi, ``Nvsim: A circuit-level performance,
  energy, and area model for emerging nonvolatile memory,'' \emph{IEEE
  Transactions on Computer-Aided Design of Integrated Circuits and Systems},
  vol.~31, no.~7, pp. 994--1007, 2012.

\bibitem{darknet13}
J.~Redmon, ``Darknet: Open source neural networks in c,''
  \url{http://pjreddie.com/darknet/}, 2013--2016.

\bibitem{che2013pannotia}
S.~Che, B.~M. Beckmann, S.~K. Reinhardt, and K.~Skadron, ``Pannotia:
  Understanding irregular gpgpu graph applications,'' in \emph{2013 IEEE
  International Symposium on Workload Characterization (IISWC)}.\hskip 1em plus
  0.5em minus 0.4em\relax IEEE, 2013, pp. 185--195.

\bibitem{che2009rodinia}
S.~Che, M.~Boyer, J.~Meng, D.~Tarjan, J.~W. Sheaffer, S.-H. Lee, and
  K.~Skadron, ``Rodinia: A benchmark suite for heterogeneous computing,'' in
  \emph{2009 IEEE international symposium on workload characterization
  (IISWC)}.\hskip 1em plus 0.5em minus 0.4em\relax Ieee, 2009, pp. 44--54.

\bibitem{leng2013gpuwattch}
J.~Leng, T.~Hetherington, A.~ElTantawy, S.~Gilani, N.~S. Kim, T.~M. Aamodt, and
  V.~J. Reddi, ``Gpuwattch: Enabling energy optimizations in gpgpus,''
  \emph{ACM SIGARCH Computer Architecture News}, vol.~41, no.~3, pp. 487--498,
  2013.

\end{thebibliography}


\begin{thebibliography}{1}

\bibitem{IEEEhowto:kopka}
H.~Kopka and P.~W. Daly, \emph{A Guide to \LaTeX}, 3rd~ed.\hskip 1em plus
  0.5em minus 0.4em\relax Harlow, England: Addison-Wesley, 1999.

\end{thebibliography}

\begin{IEEEbiography}[{\includegraphics[width=0.8in,height=1in,clip,keepaspectratio]{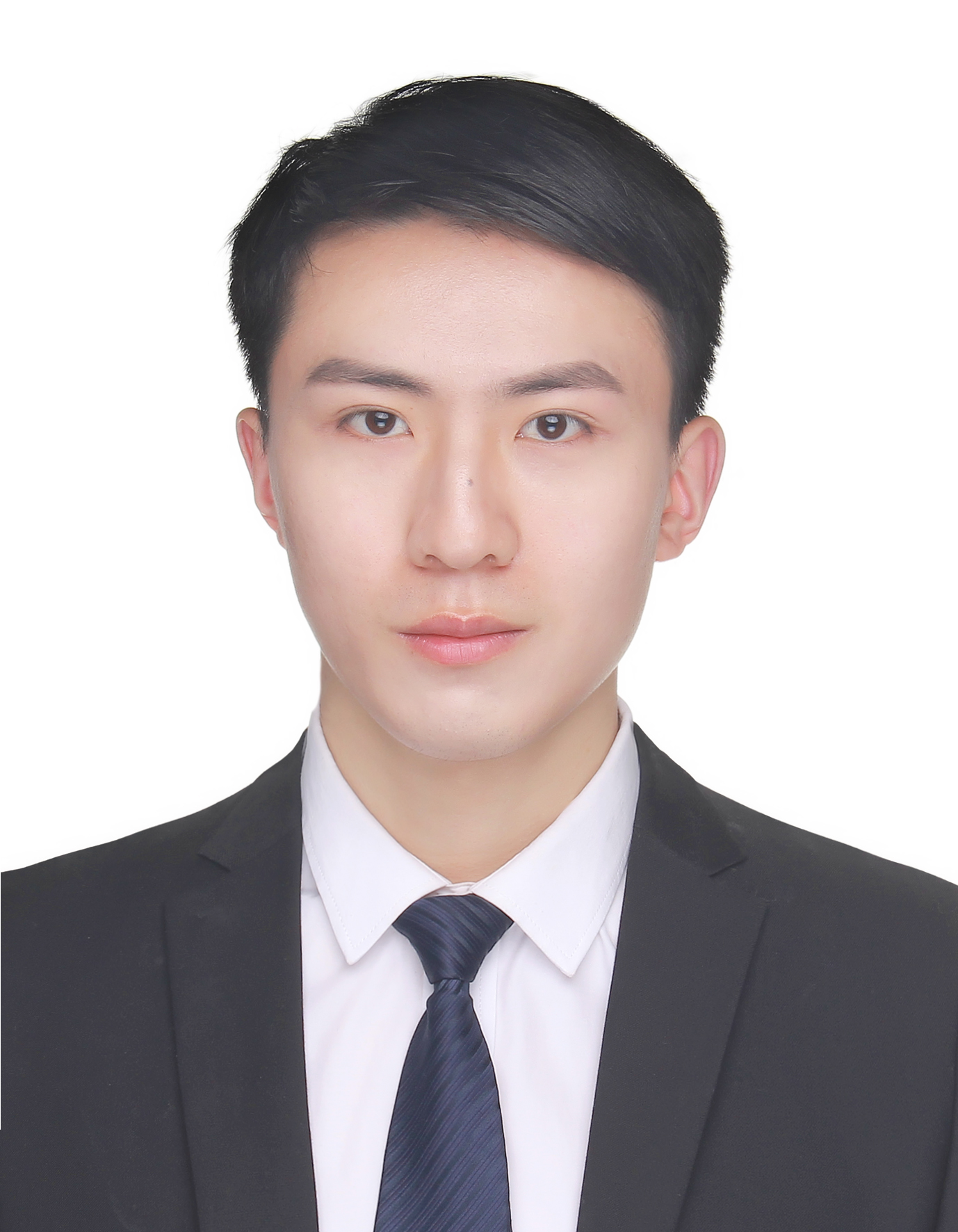}}]{Wei Zhao} received the Ph.D degree in computer architecture from Huazhong University of Science and Technology. Now, he is an engineer in Alibaba Cloud. His research interests include network filesystems, emerging non-volatile memory (NVM), 
	storage and compute micro-architecture. He publishes several papers in international conferences and journals including ICCAD, DATE, MSST, IEEE TCAD, etc.
\vspace{-40pt}	
\end{IEEEbiography}

\begin{IEEEbiography}[{\includegraphics[width=0.8in,height=1in,clip,keepaspectratio]{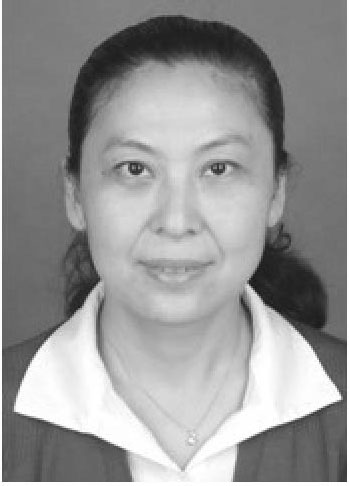}}]{Dan Feng }(IEEE Fellow) received the BE, ME, and Ph.D. degrees
	in computer science and technology from
	the Huazhong University of Science and Technology
	(HUST), China, in 1991, 1994, and 1997,
	respectively. She is a professor and vice dean of
	the School of Computer Science and Technology,
	HUST. Her research interests include computer architecture,
	massive storage systems, and parallel
	file systems. She has more than 80 publications to
	her credit in journals and international conferences,
	including IEEE TPDS, JCST, USENIX ATC, FAST, ISCA,
	ICDCS, HPDC, SC, DAC, ICS, and ICPP.
	\vspace{-30pt}	
\end{IEEEbiography}

\begin{IEEEbiography}[{\includegraphics[width=0.8in,height=1in,clip,keepaspectratio]{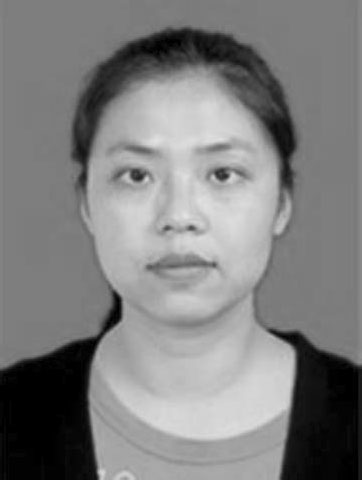}}]{Wei Tong} obtained her BE, ME, and Ph.D. degrees
	from Huazhong University of Science and Technology
	(HUST), China, respectively in 1999, 2002,
	and 2011. She is an associate professor in Wuhan
	National Laboratory for Optoelectronics, HUST. Her
	present research interests include computer architecture,
	non-volatile memory \& storage, and software-defined
	storage. She has more than 20 publications
	in journals and international conferences including
	IEEE TC, IEEE TCAD, ACM TACO, ISCA, DAC, DATE, ICCAD, ICCD, ICPP, MSST, LCTES.
	
	\vspace{-30pt}	
\end{IEEEbiography}

\begin{IEEEbiography}[{\includegraphics[width=1in,height=1.25in,clip,keepaspectratio]{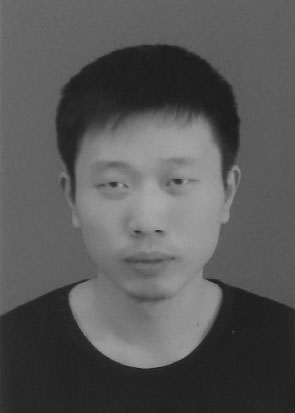}}]{Xueliang Wei} received the BE degree in computer science and technology from the Huazhong University of Science and Technology (HUST), China, in 2015. He is currently working toward the Ph.D. degree at HUST, China. His current research interests include non-volatile memories persistent memories, crash consistency, and memory security. He publishes several papers in international conferences and journals including ISCA, ICPP, TC, TOS, etc.
\end{IEEEbiography}

\begin{IEEEbiography}[{\includegraphics[width=0.8in,height=1in,clip,keepaspectratio]{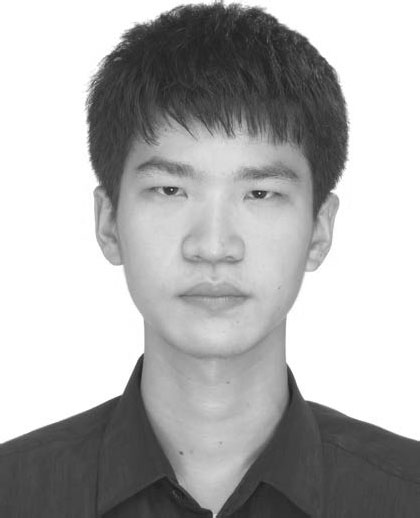}}]{Bing Wu} received the Ph.D. degree in computer
	science and technology from the Huazhong University
	of Science and Technology (HUST), China, in
	2022. He is currently a postdoc in computer
	architecture with Wuhan National Laboratory for
	Optoelectronics, HUST, China. His research interests
	include non-volatile memory (NVM) and NVMbased
	operating system. He publishes several papers
	in journals and international conferences including
	IEEE TCAD, ICCAD, ICCD, etc.
	\vspace{-40pt}
\end{IEEEbiography}

%\newpage


%\vfill


\end{document}